\documentclass[twocolumn,showpacs,preprintnumbers,amsmath,amssymb]{revtex4}
\usepackage{amsmath,amsfonts,latexsym,amssymb,graphics,epsfig,subfigure,color,makeidx}
\usepackage{xcolor}
\usepackage[colorlinks,
            linkcolor=blue,
            anchorcolor=blue,
            citecolor=green
            ]{hyperref}

\begin{document}

\title{Innermost stable circular orbit of spinning particle in charged spinning black hole background}

\author{Yu-Peng Zhang\footnote{zhangyupeng14@lzu.edu.cn},
       Shao-Wen Wei\footnote{weishw@lzu.edu.cn},
       Wen-Di Guo\footnote{guowd14@lzu.edu.cn},
       Tao-Tao Sui\footnote{suitt14@lzu.edu.cn},
       Yu-Xiao Liu \footnote{liuyx@lzu.edu.cn, corresponding author}
}
\affiliation{Institute of Theoretical Physics, Lanzhou University, Lanzhou 730000, China}

\begin{abstract}
In this paper we investigate the innermost stable circular orbit (ISCO) for a classical spinning test particle in the background of Kerr-Newman black hole. It is shown that the orbit of the spinning particle is related to the spin of the test particle. The motion of the spinning test particle will be superluminal if its spin is too large. We give an additional condition by considering the superluminal constraint for the ISCO in the black hole backgrounds. We obtain numerically the relations between the ISCO and the properties of the black holes and the test particle. It is found that the radius of the ISCO for a spinning test particle is smaller than that of a non-spinning test particle in the black hole backgrounds.
\end{abstract}

\pacs{ 04.50.-h, 11.27.+d}

\maketitle

\section{Introduction}\label{scheme1}

In the Newtonian gravitational theory we know that a massive particle can orbit around a central celestial body in the gravitational field generated by the central celestial body. Coincidentally, in general relativity a massless or massive particle can also orbit around a central celestial body and the properties of the central celestial body will affect the motion of the particle orbiting it. In Ref. \cite{Kaplan} Kaplan firstly investigated a non-spinning massive test particle orbiting a Schwarzschild black hole and found that there exists a stable circular orbit with a minimal radius $3r_h$, where the $r_h$ is the radius of the horizon of the Schwarzschild black hole. This orbit is the innermost stable circular orbit (ISCO)\footnote{For more detailed description, see the book \cite{Landau}.}. It is well known that when a black hole has spin and charge, the motion of the test particle will change, so the ISCO of the test particle will depend on both the spin and charge of the black hole. The ISCOs in different black hole backgrounds were investigated systematically in Refs. \cite{Suzuki:1997by,Shibata:1998ih,Zdunik:2000qn,Baumgarte:2001ab,Grandclement:2001ed,Miller:2003vc,Marronetti:2003hx,GondekRosinska:2004eh,Campanelli:2006gf,Shahrear:2007zz,Cabanac:2009yz,Abdujabbarov:2009az,Hadar:2011vj,Akcay:2012ea,Hod:2013vi,Chakraborty:2013kza,Hod:2014tpa,Isoyama:2014mja,Zahrani:2014rqa,Zaslavskii:2014mqa,Delsate:2015ina,Ruangsri:2015cvg,Chartas:2016ckd,Harms:2016ctx,Lukes-Gerakopoulos:2017vkj,Chaverri-Miranda:2017gxq}.

We known that the motion of a test particle should be geodesic.
When the reaction of the test particle is considered, the corresponding motion will not be geodesic any more \cite{Warburton2010,Isoyama:2014mja}. In addition to the geodesic deviation resulted by the reaction of the test particle, the spin of the test particle can also lead to a geodesic deviation for the motion of the test particle \cite{Bardeen1972,Hanson1974}. Therefore, the ISCO of a spinning test particle will also be affected by the spin of the test particle. So, it is necessary to investigate the ISCO of the spinning test particle in black hole backgrounds. The ISCO of a spinning test particle in the Schwarzschild and Kerr spacetimes was firstly investigated numerically in Ref. \cite{Suzuki:1997by}. In Ref. \cite{Jefremov:2015gza}, Jefremov, Tsupko, and Bisnovatyi-Kogan numerically investigated the ISCO of the spinning test particle in the Schwarzschild and Kerr spacetimes and gave the approximate analytic solutions of the ISCO for the particle with a small spin.

The equations of motion for a spinning particle in curved spacetime were obtained in Refs.~\cite{Mathisson,Papapetrou1951a,Papapetrou1951c,Dixon,phdthesis,Hojman1977,Zalaquett2014}. For the motion of a spinning test particle in curved spacetime, the corresponding velocity vector $u^\mu$ and the canonical momentum vector $P^\mu$ are not parallel \cite{phdthesis,Armaza2016,Hojman2013}. The canonical momentum vector $P^\mu$ keeps timelike along the trajectory and satisfies $P^\mu P_\mu=-m^2$ while the velocity vector $u^\mu$ might transform to be spacelike from timelike \cite{phdthesis,Armaza2016,Hojman2013} if the spin of the test particle is too large. In Refs.~\cite{Suzuki:1997by,Jefremov:2015gza}, the authors did not consider the superluminal constraint. In order to investigate more accurately the ISCO of a test particle with arbitrary spin $s$ in black hole background, we should add the superluminal constraint. In this paper, we will give a new condition by considering the superluminal constraint for the ISCO and numerically investigate it in Kerr-Newman (KN) black hole background with the superluminal constraint.
We find that the radius of the ISCO for a spinning test particle is smaller than that of a non-spinning test particle in KN black hole background, which is consistent with the result obtained in Refs.~\cite{Suzuki:1997by,Jefremov:2015gza}.  We also investigate how the ISCO of a spinning test particle is affected by the properties of the black hole and the spin of the test particle with the additional superluminal constraint.

Our paper is organized as follows. In Sec. \ref{scheme1} we review the equations of motion for a spinning test particle in curved spacetime and obtain the corresponding four-momentum and four-velocity in KN black hole background. In Sec. \ref{scheme2} we give a new condition for solving the ISCO of the spinning test particle with superluminal constraint in black hole background, and we also investigate how the characters of the ISCO for the spinning test particle are affected by the particle's spin $s$ and the black hole charge and spin. Finally, a brief summary and conclusion are given in Sec. \ref{Conclusion}.

\section{Motion of a spinning test particle in Kerr-Newman black hole background}{\label{scheme1}}

In this section, we review the equations of motion of a spinning test particle in curved spacetime. The effect of the spin of a test particle on its motion was first derived by considering that the test particle's spin is coupled with curvature \cite{Papapetrou1951a,Papapetrou1951c}, and the equations of motion can be derived with several methods \cite{Mathisson,Dixon,phdthesis,Hojman2013,Faye:2006gx}. Here, we use the Lagrangian to derive the equations of motion for a spinning test particle based on Refs. \cite{phdthesis,Hojman2013}. The position and orientation of a spinning test particle can be represented by the coordinate $x^\mu$ and the orthonormal tetrad $e_{(\alpha)}^\mu$, respectively. The tetrad $e_{(\alpha)}^\mu$ satisfies the relation $g^{\mu\nu}=e^{\mu}_{(\alpha)}e^{\nu}_{(\beta)}\eta^{(\alpha\beta)}$. We define the four-velocity of the spinning test particle as follows
    \begin{equation}
    u^{\mu}\equiv\frac{dx^\mu}{d\lambda},\label{velocity}
    \end{equation}
where $\lambda$ is the affine parameter. For the spinning test particle the corresponding angular velocity tensor $\sigma^{\mu\nu}$ is defined as
    \begin{equation}
    \sigma^{\mu\nu}\equiv\eta^{(\alpha\beta)}e_{(\alpha)}^\mu\frac{D e_{(\beta)}^\nu}{D\lambda}=-\sigma^{\nu\mu},\label{angularvelocity}
    \end{equation}
where $\frac{D e_{(\beta)}^\nu}{D \lambda}$ is the covariant derivative of the tetrad and has the form
    \begin{equation}
    \frac{D e_{(\beta)}^\nu}{D \lambda}\equiv\frac{de_{(\beta)}^\nu}{d\lambda}+\Gamma^\nu_{\rho\tau}e_{(\beta)}^\rho u^\tau.\label{covatiantderivative}
    \end{equation}

The Lagrangian $\mathcal{L}$ that describes the spinning test particle in curved spacetime can be constructed in terms of invariant quantities. There are four independent invariants \cite{phdthesis,Hojman1977}:
\begin{eqnarray}
a_1&=&u^\mu u_\mu,\nonumber\\
a_2&=&\sigma^{\mu\nu}\sigma_{\mu\nu}=-\texttt{tr}(\sigma^2),\nonumber\\ a_3&=&u_\alpha\sigma^{\alpha\beta}\sigma_{\beta\gamma}u^\gamma,\nonumber\\ a_4&=&g_{\mu\nu}g_{\rho\tau}g_{\alpha\beta}g_{\gamma\delta}\sigma^{\delta\mu}\sigma^{\nu\rho}
\sigma^{\tau\alpha}\sigma^{\beta\gamma}.
\end{eqnarray}
Then the final equations of motion for the spinning particle can be derived by using $\mathcal{L}=\mathcal{L}(a_1,a_2,a_3,a_4)$ \cite{phdthesis} as follows
    \begin{eqnarray}
    \frac{D P^{\mu}}{D \lambda} &=& -\frac{1}{2}R^\mu_{\nu\alpha\beta}u^\nu S^{\alpha\beta},\label{equationmotion1}\\
    \frac{D S^{\mu\nu}}{D \lambda} &=& S^{\mu\lambda}\sigma_\lambda^\nu-\sigma^{\mu\lambda}S^\nu_\lambda=P^\mu u^\nu-u^\mu P^\nu,\label{equationmotion2}
    \end{eqnarray}
where $P_\mu$ and $S_{\mu\nu}$ are the conjugate momentum vector and spin tensor, respectively, and they are defined by
    \begin{equation}
    P_\mu\equiv\frac{\partial \mathcal{L}}{\partial u^\mu},~~~~~~S_{\mu\nu}\equiv\frac{\partial \mathcal{L}}{\partial \sigma^{\mu\nu}}=-S_{\nu\mu}.
    \end{equation}
Obviously, the motion of a spinning test particle does not follow the geodesic. Here, we should note that for the spinning test particle the canonical momentum $P^\mu$ satisfies $P^\mu P_\mu=-m^2$, which means that the canonical momentum $P^\mu$ keeps timelike along the trajectory. However, things will be different for the velocity vector $u^\mu$, which may transform from timelike to spacelike as it is not parallel to $P^\mu$ \cite{phdthesis,Armaza2016,Hojman2013}.

Next we will solve the equations of motion of the spinning test particle in the KN black hole background.  The KN black hole background can be described by the Boyer-Lindquist coordinates
    \begin{eqnarray}
    ds^2&=&-\frac{\Delta}{\rho^2}(dt-a~\sin^2\theta d\phi)^2+\frac{\rho^2}{\Delta}dr^2+\rho^2 d\theta^2\nonumber\\
      &&+\frac{\sin^2\theta}{\rho^2}[(r^2+a^2)d\phi-adt]^2,\label{metric}
    \end{eqnarray}
where the metric functions $\Delta$ and $\rho^2$ are
\begin{equation}
\Delta=r^2-2Mr+a^2+Q^2, ~~ \rho^2=r^2+a^2\cos^2\theta.
\end{equation}
Here $Q$ and $a$ are the charge and spin of the black hole, respectively. The KN black hole has outer and inner  horizons $r_{\pm}=1\pm\sqrt{1-(a^2+Q^2)}$ and we have chosen $M=1$ for simplicity. The charge and spin of the KN black hole should satisfy the constraint
    \begin{equation}
    a^2+Q^2\leq 1,\label{fermionaction}
    \end{equation}
where $``="$ corresponds to the extremal black hole with one degenerate horizon.

In this paper, we only consider the equatorial motion of the test particle with $\theta=\frac{\pi}{2}$. So the non-vanishing  components of the conjugate momentum are \cite{Hojman1977,Zhang:2016btg}
    \begin{eqnarray}
    P^t&=&\frac{m^3}{\Theta \Xi}
       \bigg[r^2 a\bar{j}(2Mr-Q^2)-\bar{e} r^6+(2a~\bar{e}-\bar{j})Q^2 r^2\bar{s}\nonumber\\
    &&-r^2a^2\bar{e}(2Mr+r^2-Q^2)
      +(\bar{j}-3a~\bar{e}) M r^3 \bar{s} \nonumber\\
    &&+a^2(a~\bar{e}-\bar{j})(Q^2-M~r)\bar{s}
    \bigg],\label{memantumpt}
    \end{eqnarray}
    \begin{eqnarray}
    P^\phi&=&\frac{m^3}{\Theta \Xi}\bigg[a \bar{e} r^2 (Q^2-2Mr)+a^2 \bar{e}(Q^2-Mr)\bar{s}+r^2Q^2\nonumber\\
    &&+a\bar{j} (Mr-Q^2)\bar{s}+r^2 (r^2-2Mr)(\bar{e}\bar{s}-\bar{j})\bigg],\label{memantumpp}
    \end{eqnarray}
    and
    \begin{eqnarray}
    (P^r)^2&=&\frac{m^6}{r^2 \Xi^2}\bigg[r^6
     \bigg(2 M  r^3 -r^2\left(\bar{j}^2+a^2-a^2\bar{e}^2+ Q^2\right)\nonumber\\
    &&+j_e^2 (2M~ r-Q^2)+(\bar{e}^2-1)r^4\bigg)\nonumber\\
    &&-2 r^4\bar{s}\left(aQ_rj_e^2-2\bar{e}j_eQ^2 r^2+3\bar{e}j_eMr^3-\bar{e}\bar{j} r^4\right)\nonumber\\
    &&-Q_r^2\Theta \bar{s}^4+r^2 \bar{s}^2a^2 Q_r\bar{e}^2 (Q_r+2 r^2)\nonumber\\
    &&+r^2 \bar{s}^2\bigg(\bar{j}^2 Q_r^2-2 a \bar{e}\bar{j}Q_r\left(Q_r+r^2\right)-a^2 Q_r2 r^2\nonumber\\
    &&-r^2 \left(Q^2+r^2-2Mr\right) \left(\bar{e}^2 r^2+2Q_r\right)\bigg)\bigg].\label{memantumpr}
    \end{eqnarray}
Here the parameters $\bar{e} \equiv \frac{e}{m}$, $\bar{s}\equiv\frac{s}{m}$, and $\bar{j}\equiv\frac{j}{m}=\frac{l}{m}+\frac{s}{m}$ are the energy, spin angular momentum, and total angular momentum per unit mass of the test particle, respectively, and  $\Theta$, $\Xi$, and $Q_r$ are defined as \cite{Zhang:2016btg}
 \begin{eqnarray}
    \Theta~ &\equiv& a^2+Q^2-2Mr+r^2,\\
    \Xi~ &\equiv& m^2 r^4+(Q^2-Mr)m^2\bar{s}^2,\\
    Q_r &\equiv& Q^2-Mr,\quad j_e\equiv\bar{j}-a\bar{e}.
    \end{eqnarray}
The velocity $u^\mu$  can be solved according to the equations of motion (\ref{equationmotion1}) and (\ref{equationmotion2}) \cite{Hojman2013}
\begin{eqnarray}
\frac{DS^{tr}}{D\lambda}&=&P^t\dot{r}-P^r\label{velocityequation1}\\
\frac{DS^{t\phi}}{D\lambda}&=&P^t\dot{\phi}-P^\phi\label{velocityequation2}.
\end{eqnarray}
The non-vanishing components of the spin tensor $S^{\mu\nu}$ in the KN black hole background are
\begin{eqnarray}
S^{r\phi}&=&-S^{\phi r}=-\frac{s P_{t}}{m r},\nonumber\\
S^{rt}&=&-S^{tr}=-S^{r\phi}\frac{P_\phi}{P_t}=s\frac{P_\phi}{mr},\label{spinnozo}\\
S^{\phi t}&=&-S^{t\phi}=S^{r\phi}\frac{P_r}{P_t}=-s\frac{P_r}{mr}.\nonumber
\end{eqnarray}
Equations \eqref{velocityequation1} and \eqref{velocityequation2} can be expressed in terms of Eq. (\ref{spinnozo}) as follows
\begin{eqnarray}
\frac{DS^{tr}}{D\lambda}&=&P^t\dot{r}-P^r\nonumber\\
&=&\frac{1}{2}\frac{s}{mr}g_{\phi \mu}R^\mu_{\nu\alpha\beta}u^\nu S^{\alpha\beta}+s\frac{P_\phi}{mr^2}\dot{r}\label{spinvelocityequation1}
\end{eqnarray}
and
\begin{eqnarray}
\frac{DS^{t\phi}}{D\lambda}&=&P^t\dot{\phi}-P^\phi\nonumber\\
&=&-\frac{1}{2}\frac{s}{mr}g_{r \mu}R^\mu_{\nu\alpha\beta}u^\nu S^{\alpha\beta}-s\frac{P_r}{mr^2}\dot{r}\label{spinvelocityequation2}.
\end{eqnarray}
Substituting the non-vanishing components of the Riemman curvature tensor of the KN black hole background into Eqs. \eqref{spinvelocityequation1} and \eqref{spinvelocityequation2}, the non-zero components of the four-velocity are \cite{Zhang:2016btg}
\begin{widetext}
\begin{eqnarray}
\dot{\phi}&=&K_2^{-1}\Bigg\{a^3 P_t s^2 (4 q^2-3 r)+[a^2+q^2+r(r-2)] \Bigg[\frac{K_1}{K_2}+m^2 P^\phi r^6+P_\phi s^2 (3 r-4 q^2)\Bigg]\nonumber\\
&&+a^2 \Big[P_\phi s^2 (3 r-4 q^2)\Big]+[q^2+r(r-2)] \Big[P_\phi s^2 (2 r-3 q^2)\Big]
  +a P_t s^2 \Big[3 q^4+4 q^2 (r-2) r+r^2 (4-3 r)\Big] \Bigg\}\label{spinvelocityp}
\end{eqnarray}
and
\begin{eqnarray}
\dot{r}&=& P^r\Delta \left[m^2 r^6 + r^2(q^2 - r)s^2\right]
   \bigg[m^2 P^t r^6 \Delta + \Big[(-P_\phi + a P_t) q^2 (4 a^3 + 3 a q^2)-m s P_\phi r^4 \Delta +a (3 P_\phi - 4 a P_t) r^3 \nonumber\\
&& +a (P_\phi - a P_t) (3 a^2 + 8 q^2) r
   + a [-4 P_\phi (1 + q^2) + a P_t (4 + 5 q^2)] r^2+ P_t q^2 r^4 - P_t r^5\Big] s^2\bigg]^{-1},\label{spinvelocityr}
\end{eqnarray}
where $v^r=\dot{r}$, $v^\phi=\dot{\phi}$, and $K_i$ ($i=1, 2$) is defined as follows
\begin{eqnarray}
K_1&=&-m P_r r^4 s \Delta \left[P_r s^2 (q^2-r) \Delta+m^2 P^r r^6\right],\\
K_2&=& m^2 P^t r^6 \Delta
       -m P_\phi r^4 s \Delta+s^2 \Big[q^2 (4 a^3+3 a q^2) (a P_t-P_\phi)+a r (3 a^2+8 q^2) (P_\phi-a P_t)\nonumber\\
&&+a r^2 \left(a P_t (5 q^2+4)-4 P_\phi (q^2+1)\right)+a r^3 (3 P_\phi-4 a P_t)+P_t q^2 r^4-P_t r^5\Big].
\end{eqnarray}
\end{widetext}
Obviously, it can be seen from Eq.~\eqref{spinvelocityr} that the radial momentum $P^r$ and redial velocity $v^r$ are parallel. So we can use the radial component $P^r$ of the four-momentum to define the effective potential for the spinning test particle.

\section{ISCO of a spinning particle in Kerr-Newman black hole background}\label{scheme2}

In this section, we will investigate the ISCO of the spinning test particle in different black hole backgrounds. As stated in the previous paper and book \cite{Kaplan,Landau}, the motion of a test particle in a central field can be solved in terms of the radial coordinate ``effective potential" in the Newtonian dynamics. And the so-called ``effective potential" method is also generalized to general relativity to solve the motion of a test particle in black hole backgrounds. We know that if a test particle satisfies the following two conditions \cite{Jefremov:2015gza}:

(a) the radial velocity of the test particle vanishes:
\begin{equation}
\frac{dr}{d\lambda}=0\label{condition1}.
\end{equation}

(b) the radial velocity should keep unchanged, which means that the acceleration of the radial velocity should be zero:
\begin{equation}
\frac{d^2r}{d\lambda^2}=0\label{condition2}.
\end{equation}
Then the corresponding trajectory of the test particle must be a stable circular orbit. As stated in Ref. \cite{Jefremov:2015gza}, we know that there is an ISCO when radius of the stable circular orbit is minimal. So the ISCO locates at the point that the maximum and minimum of the effective potential merge. It is obvious that for the ISCO the effective potential of the test particle should also satisfy
\begin{equation}
\frac{d^2V_{\text{eff}}}{dr^2}=0\label{condition3}.
\end{equation}
So we can use these three conditions \eqref{condition1}, \eqref{condition2}, and \eqref{condition3} to get the ISCO of the test particle.

\begin{figure}[!htb]
    \includegraphics[width=0.23\textwidth]{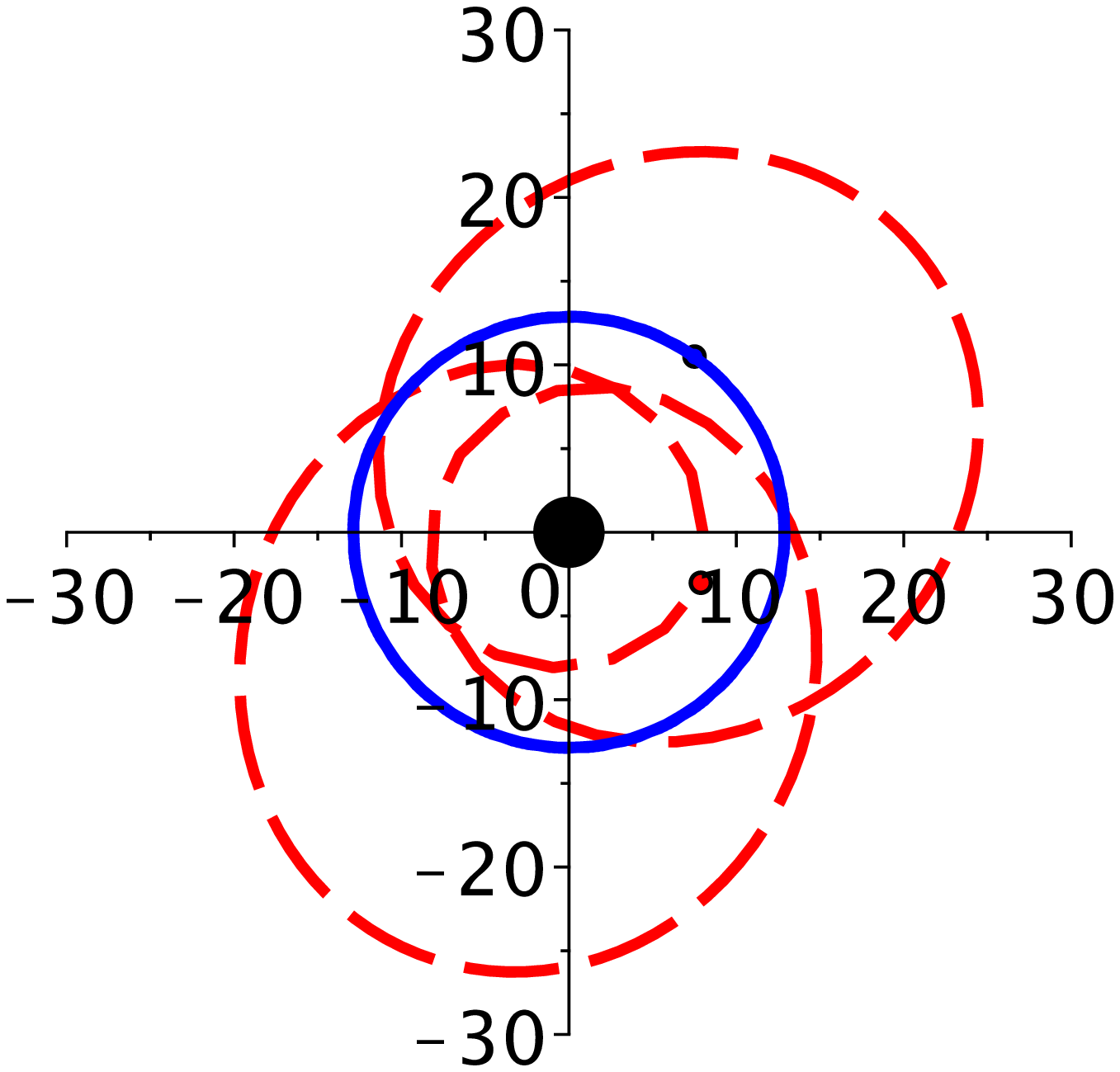}
    \includegraphics[width=0.23\textwidth]{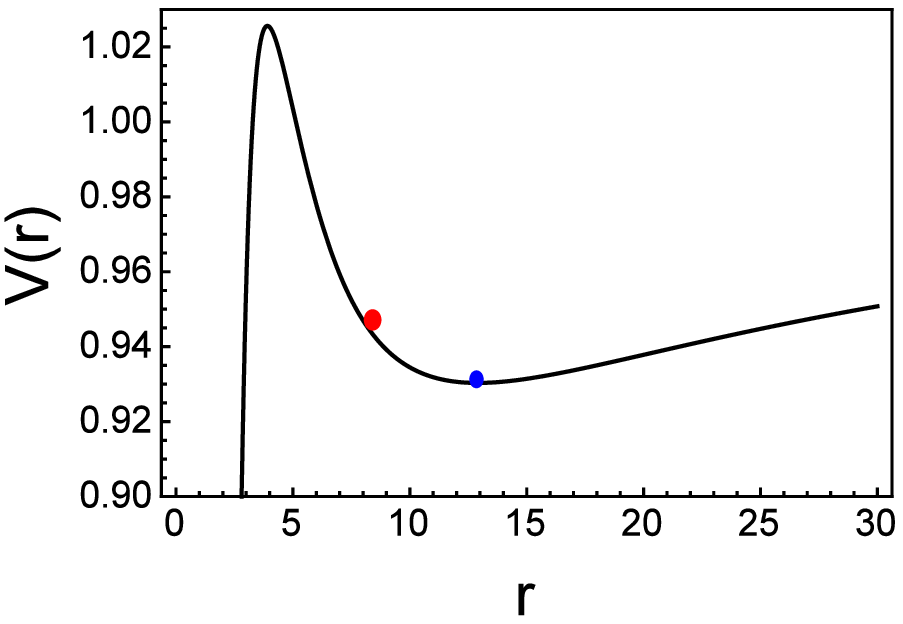}
    \vskip -4mm \caption{Plots of the orbits and effective potential for the non-spinning test particle in the Schwarzschild black hole background. The blue dot means the particle locates at the minimum value of the effective potential and the corresponding orbit is circular orbit (blue solid line), while the red dot stands for the test particle with energy $\bar{e}\approx 0.9461$ and orbital angular momentum $\bar{l}=4.1$ and the corresponding orbit (red dashed line) is not a circular orbit.}
    \label{orbiteffective}
    \end{figure}

For the Schwarzschild black hole, the corresponding effective potential of a non-spinning test particle is
\begin{equation}
V_{\text{eff}}^{\text{Schw}}=\sqrt{\left(1-\frac{2M}{r}\right)\left(1+\frac{\bar{l}^2}{r^2}\right)},\label{effectivepotentialshcw}
\end{equation}
and the parameters of the ISCO of the test particle are \cite{Kaplan}
\begin{eqnarray}
r_{\text{ISCO}}=6M,~~ \bar{l}_{\text{ISCO}}=2\sqrt{3}M,~~ \bar{e}_{\text{ISCO}}=\sqrt{\frac{8}{9}},
\end{eqnarray}
where the parameters $\bar{l}$ and $\bar{e}$ are the orbital angular momentum and energy per unit rest mass of the test particle, respectively. If the test particle moves along \textcolor[rgb]{0.00,0.00,1.00}{a} circular orbit, its energy should be the minimum value of the effective potential. For example, the corresponding orbits of the non-spinning test particle with different energies are shown in Fig.~ \ref{orbiteffective}.

For the Kerr black hole, the ISCO was given in Ref. \cite{Bardeen1972} for the extremal case with $a=M$. Due to the drag effect of the
Kerr black hole, the ISCOs with counter-rotating orbit and co-rotating orbit are different and the corresponding results are respectively
\begin{eqnarray}
r_{\text{ISCO}}=9M,~~ \bar{l}_{\text{ISCO}}=-\frac{22}{3\sqrt{3}}M,~~ \bar{e}_{\text{ISCO}}=\frac{5}{3\sqrt{3}},\label{kerrisco1}
\end{eqnarray}
and
\begin{eqnarray}
r_{\text{ISCO}}=M,~~ \bar{l}_{\text{ISCO}}=\frac{2}{\sqrt{3}}M,~~ \bar{e}_{\text{ISCO}}=\frac{1}{\sqrt{3}}.\label{kerrisco2}
\end{eqnarray}
Next, we will investigate the ISCO of the spinning test particle in the KN black hole background. In Sec. \ref{scheme1} we have solved the four-momentum and velocity of the spinning test particle by using the equations of motion \eqref{equationmotion1} and \eqref{equationmotion2}. Since the radial velocity and radial component $P^r$ of the four-momentum are parallel, we can use $P^r$ to define the effective potential of the spinning test particle in black hole background. The square of $P^r$ reads
    \begin{eqnarray}
    (P^r)^2&=&\frac{m^6}{r^2 \Xi^2}\Bigg[r^6 \bigg(2 M  r^3 -r^2\left(\bar{j}^2+a^2-a^2\bar{e}^2+ Q^2\right)\nonumber\\
    &&+j_e^2 (2M~ r-Q^2)+(\bar{e}^2-1)r^4\bigg)\nonumber\\
    &&-2 r^4\bar{s}\left(aQ_rj_e^2-2\bar{e}j_eQ^2 r^2+3\bar{e}j_eMr^3-\bar{e}\bar{j} r^4\right)\nonumber\\
    &&-Q_r^2\Theta \bar{s}^4+r^2 \bar{s}^2a^2 Q_r\bar{e}^2 (Q_r+2 r^2)\nonumber\\
    &&+r^2 \bar{s}^2\bigg(\bar{j}^2 Q_r^2-2 a \bar{e}\bar{j}Q_r\left(Q_r+r^2\right)-a^2 Q_r2 r^2\nonumber\\
    &&-r^2 \left(Q^2+r^2-2Mr\right) \left(\bar{e}^2 r^2+2Q_r\right)\bigg)\Bigg],\label{memantumpr1}
    \end{eqnarray}
which can be decomposed as follows
    \begin{eqnarray}
    (P^r)^2&=&\frac{m^6}{r^2 \Xi^2}(\alpha e^2+\beta e+\gamma)\nonumber\\
    &=&\frac{m^6}{r^2 \Xi^2}\left(e-\frac{-\beta+\sqrt{\beta^2-4\alpha\gamma}}{2\alpha}\right)\nonumber\\
    &&\times \left(e+\frac{\beta+\sqrt{\beta^2-4\alpha\gamma}}{2\alpha}\right),
    \end{eqnarray}
where the functions $\alpha$, $\beta$, and $\gamma$ are
\begin{eqnarray}
\alpha &=&m^2 r^6 \left(a^2 (2Mr+r^2-Q^2)+r^4\right)\nonumber\\
&&+2amr^4 s \left(a^2 \left(M r-Q^2\right)+r^2 \left(3 M r-2 Q^2\right)\right)\nonumber\\
&&+r^2 s^2 \bigg[a^2 \left(Q^2-M r\right) \left(Q^2-r (M+2 r)\right)\nonumber\\
&&-r^4 \left(r (r-2 M)+Q^2\right)\bigg],
\end{eqnarray}
\begin{eqnarray}
\beta &=&2 j r^2 \Big[2a^2mr^2s(Q^2-M r)+am^2r^4(Q^2-2Mr)\nonumber\\
&&-as^2(Q^2-Mr)(Q^2-r(M+r))\nonumber\\
&&+m r^4s(r(r-3M)+2Q^2)\Big],
\end{eqnarray}
and
\begin{eqnarray}
\gamma &=&j^2 r^2 \Big[2 a m r^2 s \left(M r-Q^2\right)\nonumber\\
&&-m^2 r^4 \left(-2 M r+Q^2+r^2\right)+s^2 \left(Q^2-M r\right)^2\Big]\nonumber\\
&&-\Delta \left(s^2 \left(Q^2-M r\right)+m^2 r^4\right)^2.
\end{eqnarray}
The effective potential of the spinning test particle in the KN black hole background is
\begin{equation}
V_{\text{eff}}^{\text{spin}}=\frac{-\beta+\sqrt{\beta^2-4\alpha\gamma}}{2\alpha}\label{spinningeffective}.
\end{equation}

    \begin{figure}[!htb]
    \centering
    \subfigure[~$a=0, Q=0, l=4$ ]{
    \includegraphics[width=0.23\textwidth]{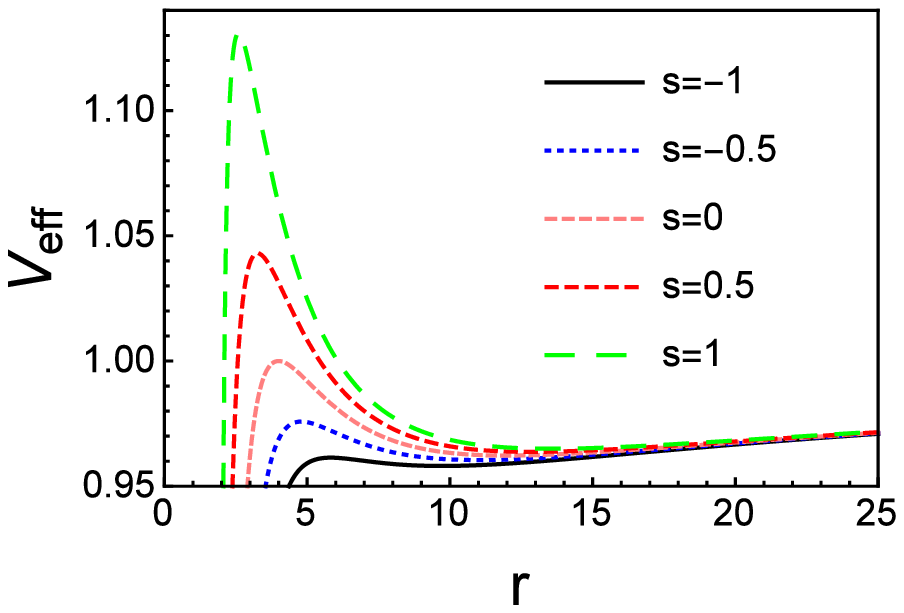}}
    \subfigure[~$a=1, Q=0, l=2$ ]{
    \includegraphics[width=0.23\textwidth]{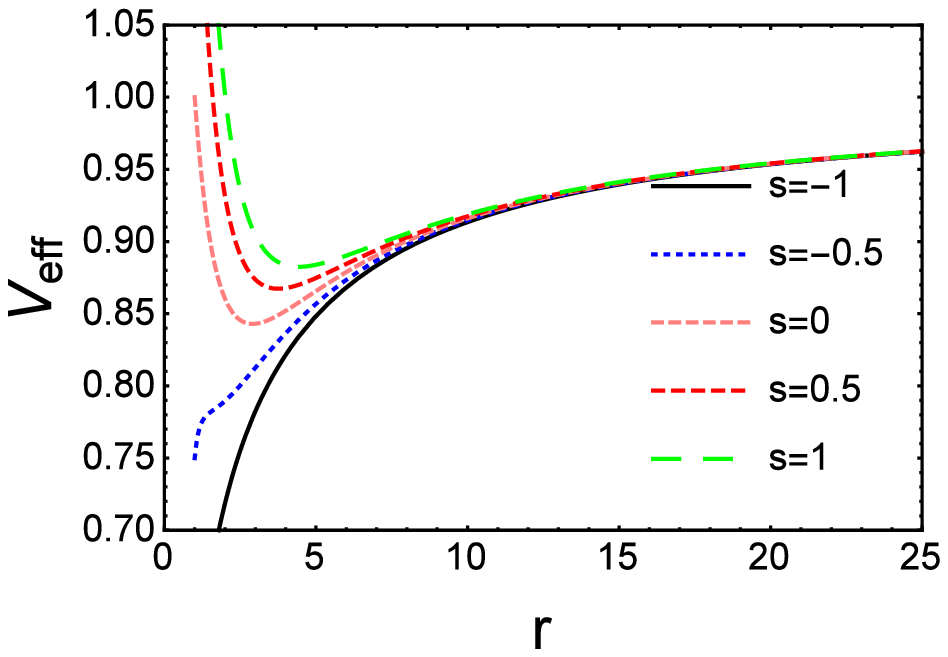}}
    \subfigure[~$a=0, Q=1, l=3.25$ ]{
    \includegraphics[width=0.23\textwidth]{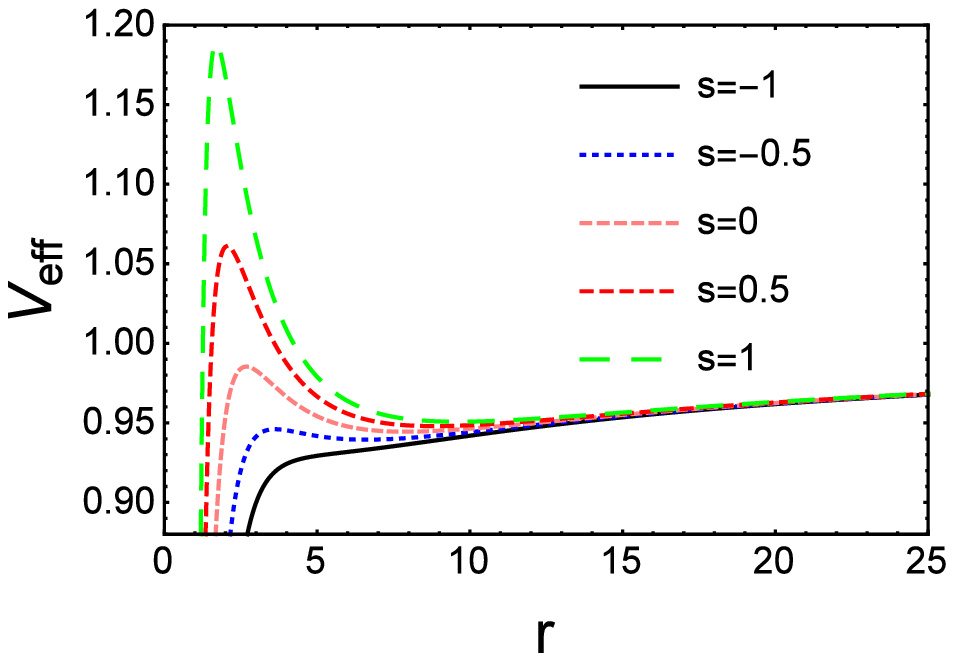}}
    \subfigure[~$a=0.8, Q=0.6, l=1$ ]{
    \includegraphics[width=0.23\textwidth]{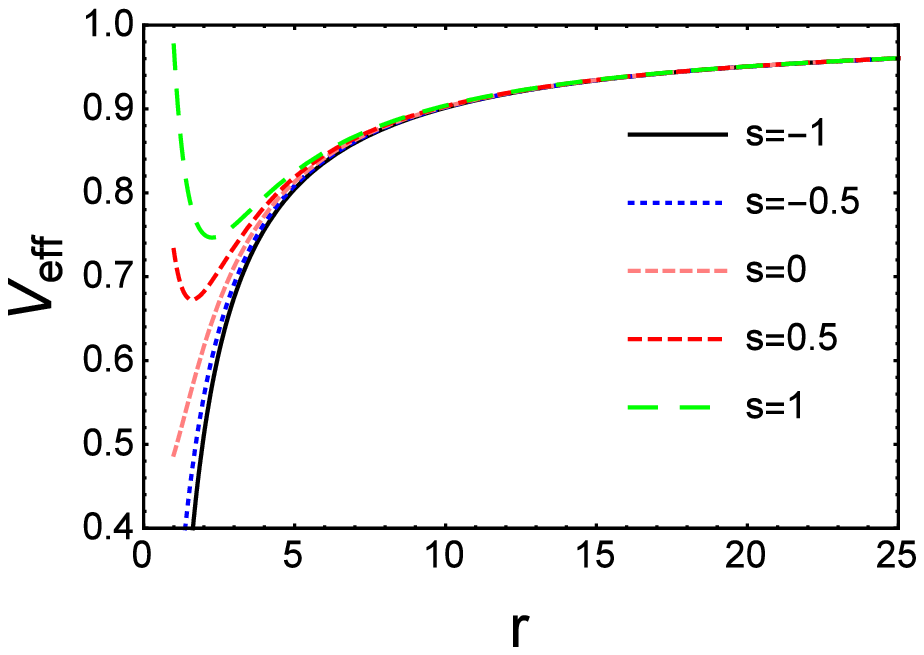}}
    \vskip -4mm \caption{Plots of the effective potential \eqref{spinningeffective} for the spinning test particle with different spins.}
    \label{effectivepotential}
    \end{figure}

For the case of $s=0$, our result \eqref{spinningeffective} can reduce to that of the KN black hole. Plots of the effective $V_{\text{eff}}^{\text{spin}}$ with different spins are shown in Fig.~\ref{effectivepotential}.

\begin{figure}[!htb]
    \centering
    \subfigure[~$a=0, Q=0$ ]{
    \includegraphics[width=0.23\textwidth]{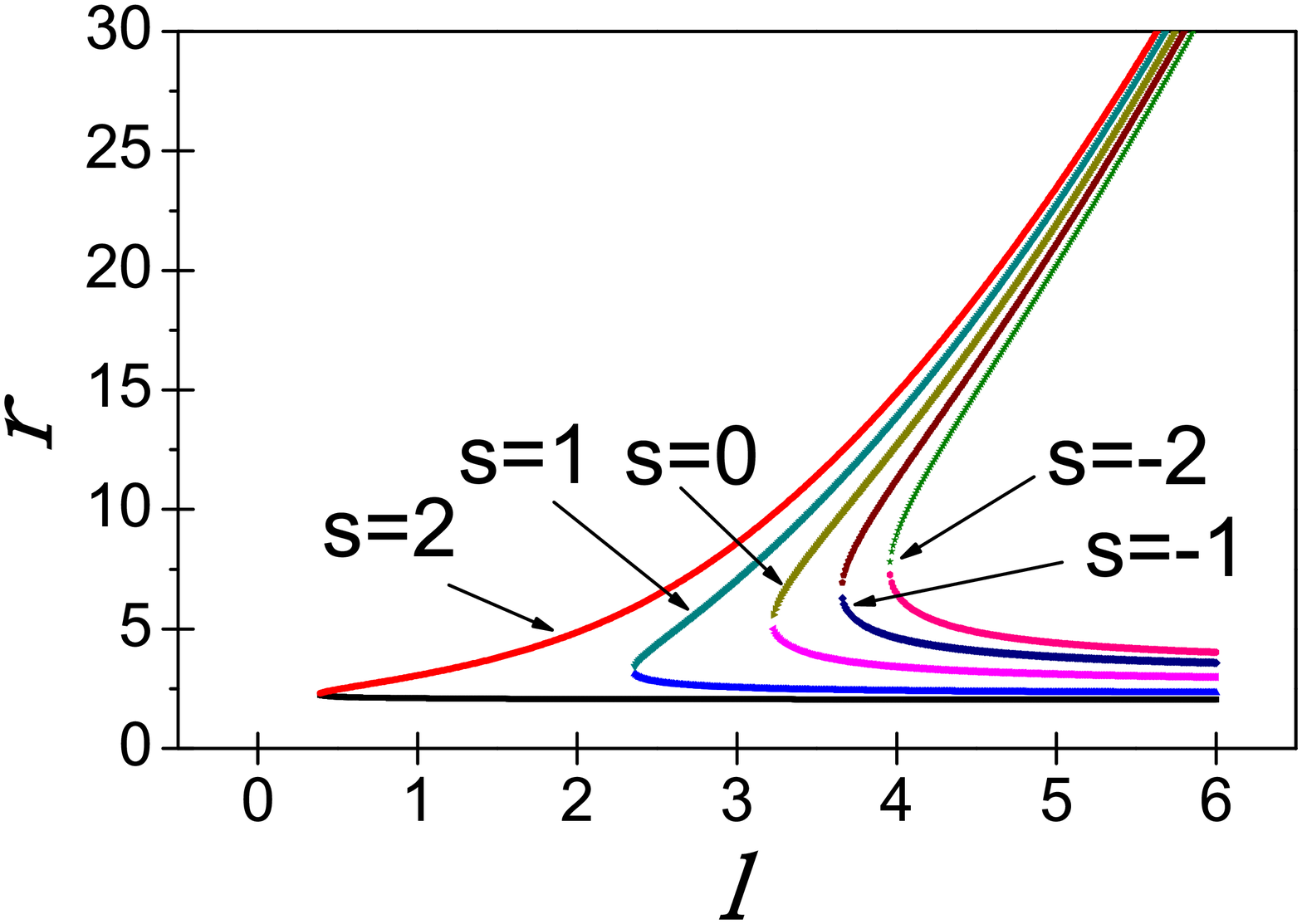}}
    \subfigure[~$a=0, Q=0.5$]{
    \includegraphics[width=0.23\textwidth]{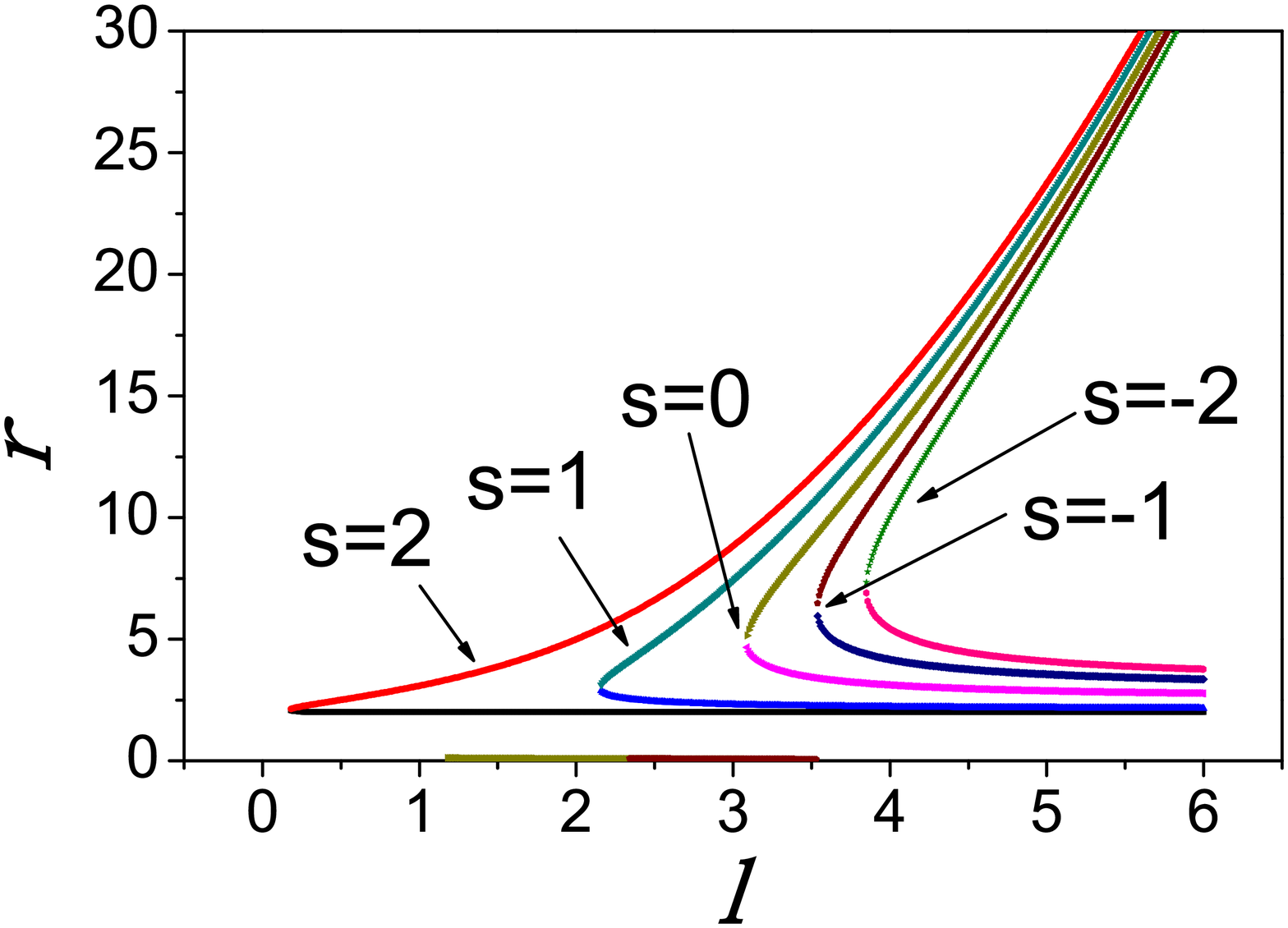}}
    \subfigure[~$a=0.5, Q=0$]{
    \includegraphics[width=0.23\textwidth]{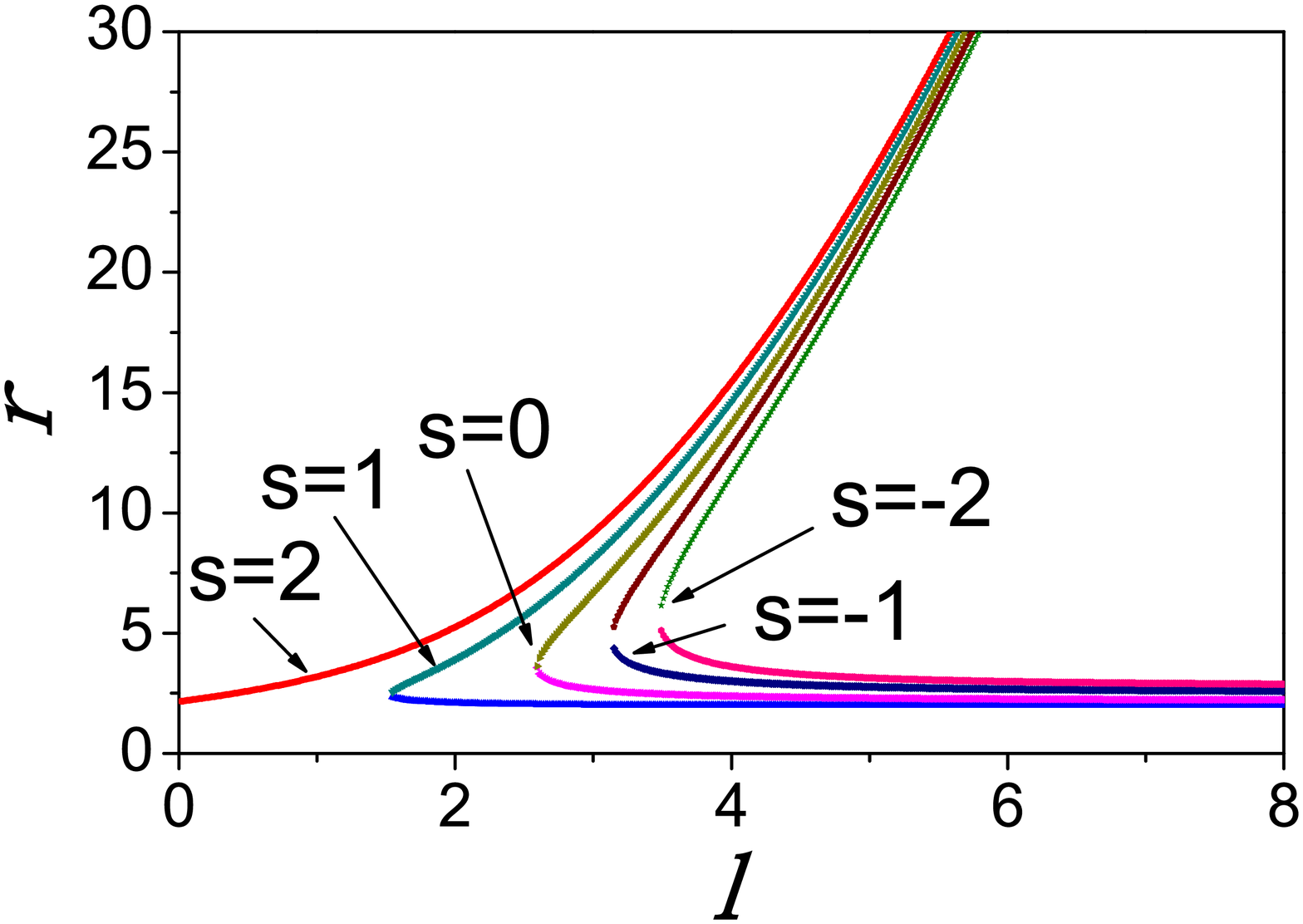}}
    \subfigure[~$a=0.5, Q=0$]{
    \includegraphics[width=0.23\textwidth]{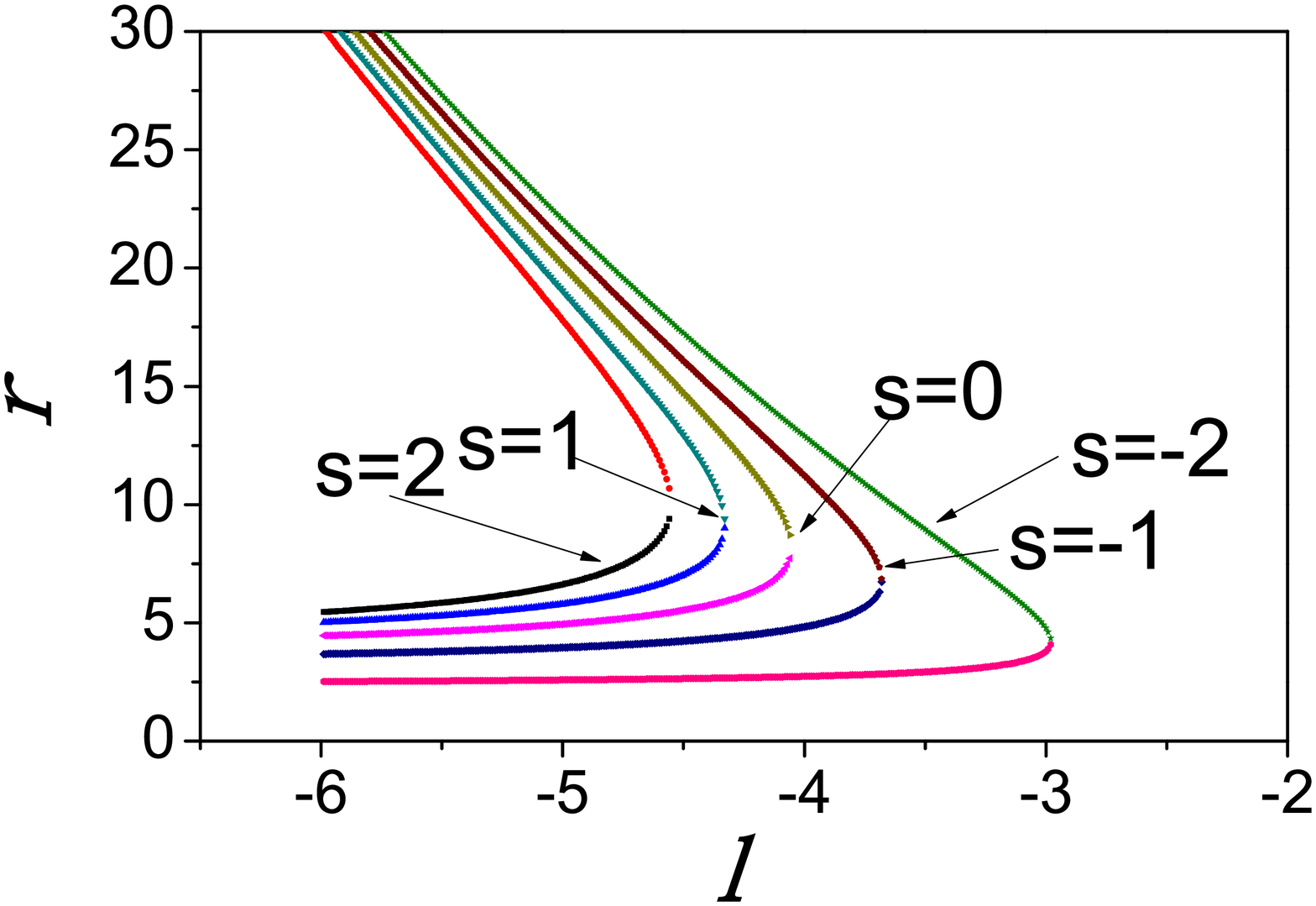}}
    \subfigure[~$a=0.5, Q=0.5$]{
    \includegraphics[width=0.23\textwidth]{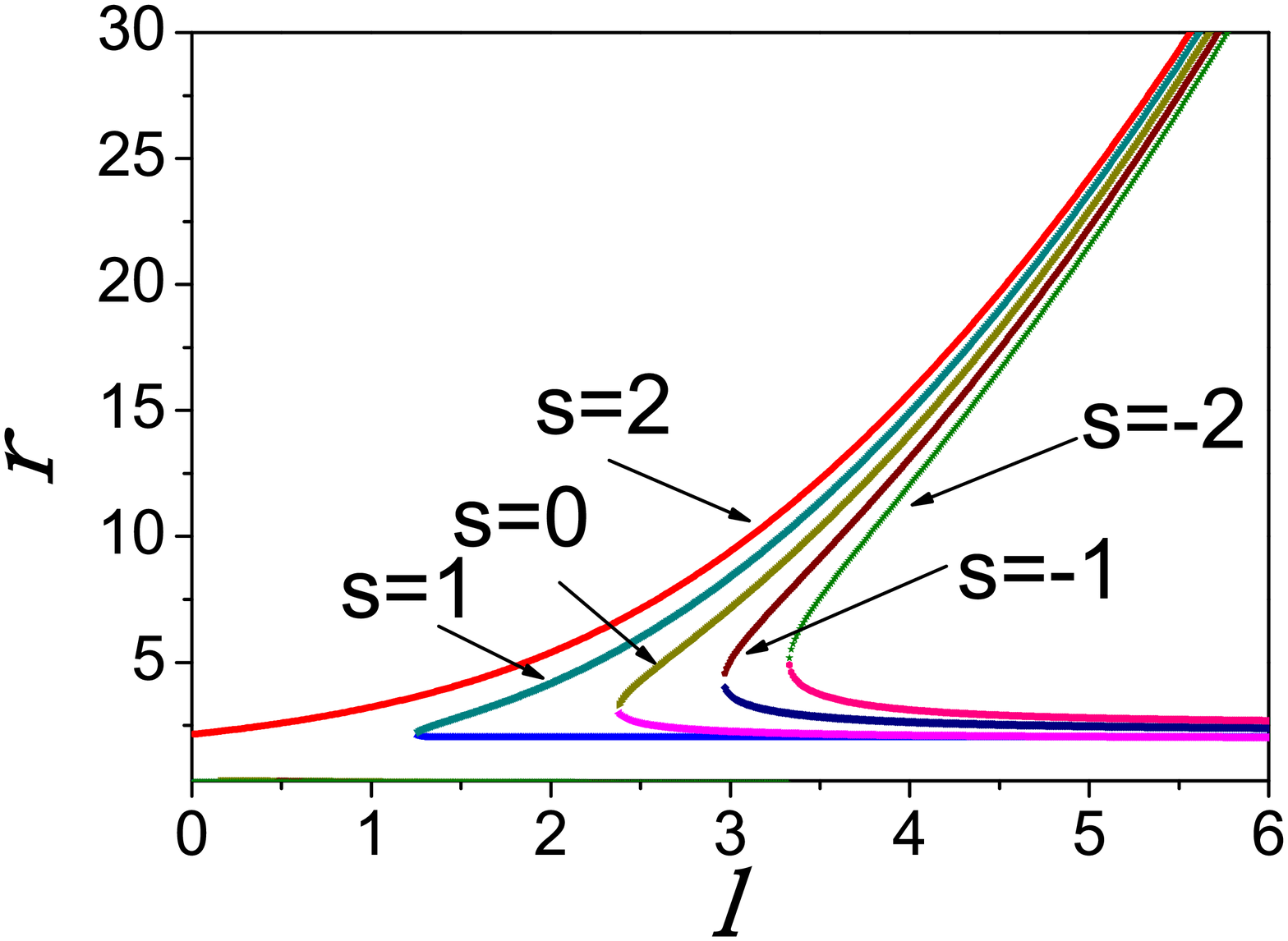}}
    \subfigure[~$a=0.5, Q=0.5$]{
    \includegraphics[width=0.23\textwidth]{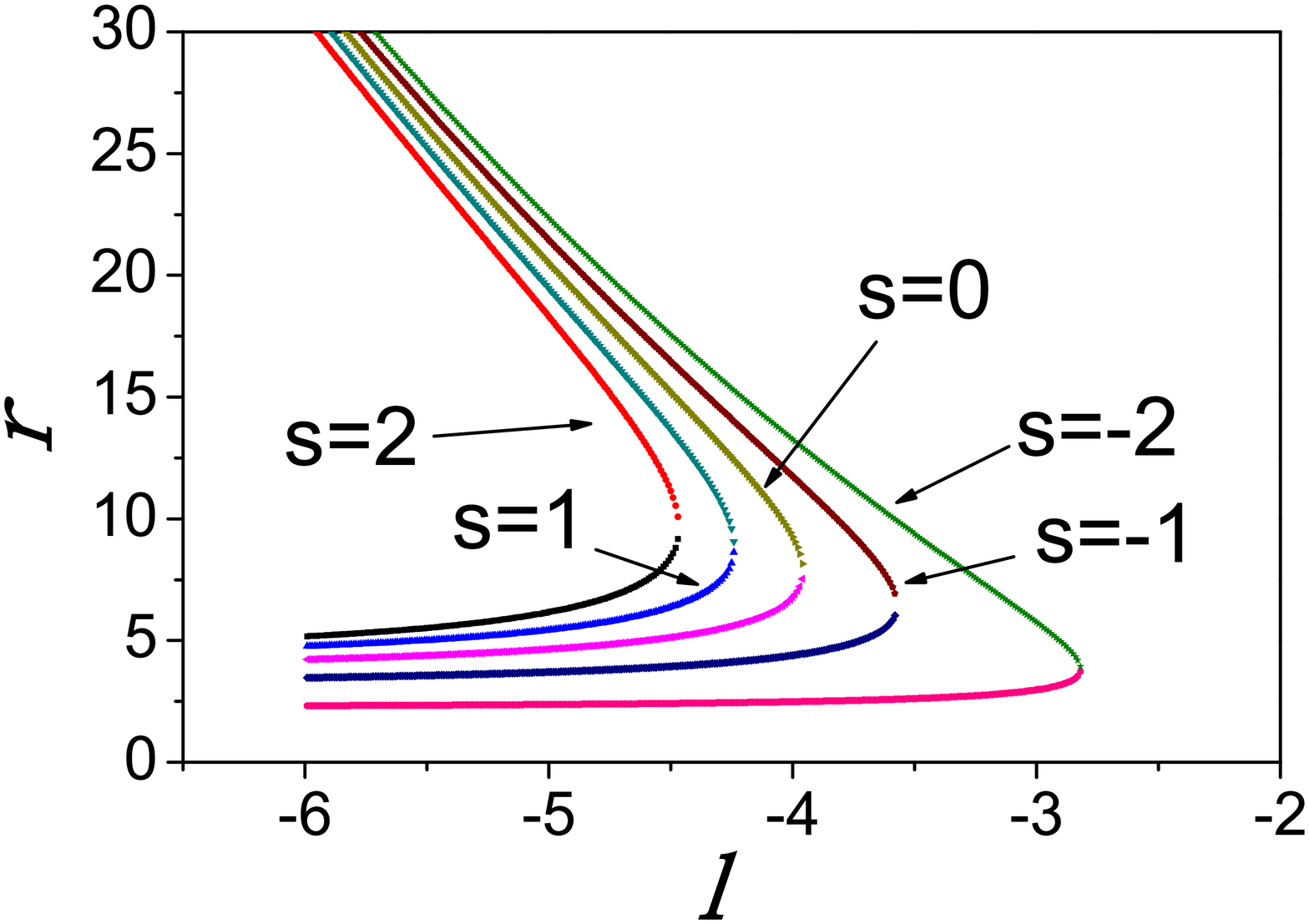}}
    \vskip -4mm \caption{The relation between circular orbit radius $r$ and orbital angular momentum $l$ for different values of the spin $s$, where the upper curves stand for the stable circular orbits while the lower curves stand for the unstable circular orbits. Here the subfigures (c) and (e) describe the co-rotating orbits and subfigures (d) and (f) describe counter-rotating orbits.}
    \label{circularorbitradius}
    \end{figure}

Note that the effective potential of the spinning test particle is dependent on the spin of the test particle.
We know that there are two extreme points in the effective potential. The orbits of the test particle corresponding to the two extreme points are circular, and one is unstable while the other is stable. We give the relation of the circular orbit radius $r$ and the orbital angular momentum $l$ with different values of the spin $s$ in Fig.~\ref{circularorbitradius}. The point that the upper and lower curves intersect defines the ISCO of the spinning test particle. It is evident that the radius of the ISCO decreases (increases) with the spin $s$ when the direction of the angular momentum of the test is the same as (the opposite of) that of the black hole.

We can solve the ISCO of the spinning test particle in terms of the three conditions \eqref{condition1}, \eqref{condition2}, and \eqref{condition3}. Here we note that the four-velocity and four-momentum are not parallel \cite{phdthesis,Armaza2016,Zhang:2016btg,Hojman2013} and the velocity may transform from timelike to spacelike, which means that the ISCO of the spinning test particle may be unphysical. So we should add the superluminal constraint
\begin{eqnarray}
\frac{u^\mu u_\mu}{(u^t)^2} =
  \frac{g_{tt}}{c^2}
  + g_{rr}\Big(\frac{\dot{r}}{c}\Big)^2
  + g_{\phi\phi}\Big(\frac{\dot{\phi}}{c}\Big)^2
  + 2g_{\phi t}\dot{\phi}<0, \label{velocitysquare}
\end{eqnarray}
which ensures that the motion of the spinning test particle in circular orbit is subluminal. In Refs. \cite{Ruangsri:2015cvg,Jefremov:2015gza}, the authors obtained the analytical corrections to the ISCO for the spinning test particle with small-spin linear approach for the Schwarzschild and Kerr black holes, and numerically investigated the effect of the spin on the ISCO without the superluminal constraint \eqref{velocitysquare}.

\begin{figure}[!htb]
    \includegraphics[width=0.35\textwidth]{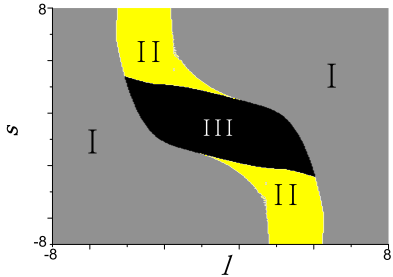}
    \vskip -4mm \caption{Plot of the region that the spinning test particle has a circular orbit in the $(s-l)$ parameter space. Region II (yellow region) stands for that the motion of the spinning test particle in the circular orbit is superluminal and unphysical, region III (black region) stands for that the spinning test particle does not have a circular orbit, region I (gray region) stands for that the spinning test particle can have a physical circular orbit.}
    \label{circularorbitregion}
    \end{figure}

In summary, if we want to solve the physical ISCO of the spinning test particle, we should use the four conditions \eqref{condition1}, \eqref{condition2}, \eqref{condition3}, and \eqref{velocitysquare}. Firstly, we numerically give the region that whether the spinning test particle has a timelike circular orbit in the $(s-l)$ parameter space. For simplicity, we only give the result for the Schwarzschild black hole in Fig.~\ref{circularorbitregion}. Obviously, some circular orbits in the parameter space are spacelike and unphysical. So it is necessary to consider the constraint \eqref{velocitysquare} for the ISCO of the spinning test particle.

\begin{figure}[!htb]
    \includegraphics[width=0.35\textwidth]{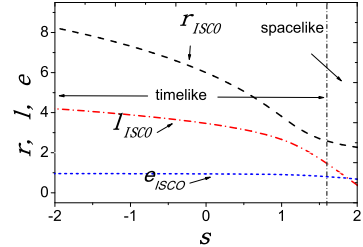}
    \vskip -4mm \caption{Plot of the ISCO parameters of the spinning test particle as functions of the spin $s$ in the Schwarzschild black hole background. The left side of the vertical line stands for that the ISCO is timelike and physical, while the right side stands for that the orbit is spacelike and unphysical.}
    \label{schiscoparameter}
    \end{figure}

To proceed, we numerically investigate the effects of the spin of the test particle on the ISCO in different black hole backgrounds with the superluminal constraint \eqref{velocitysquare} in detail. Firstly, we give the complete numerical results for the ISCO of the spinning test particle in the Schwarzschild black hole background. The corresponding numerical results are shown in Fig.~\ref{schiscoparameter}. It can be seen that the ISCO parameters of the spinning test particle will decrease with the spin $s$. This is the same as the result given in Ref. \cite{Ruangsri:2015cvg}, but the superluminal constraint was not considered there. Here some of these orbits are superluminal and unphysical.
Now the spin of the test particle is one of the parameters of the ISCO. The physical ISCO parameters of the spinning test particle in the Schwarzschild black hole background are calculated as follows
\begin{eqnarray}
\bar{s}^{\text{Sch}}_{\text{ISCO}}&\thickapprox&1.6510M,~~~r^{\text{Sch}}_{\text{ISCO}}\thickapprox2.5308M,\nonumber\\
\bar{e}^{\text{Sch}}_{\text{ISCO}}&\thickapprox&0.7896,~~~\,\,\,~~\bar{l}^{\text{Sch}}_{\text{ISCO}}\thickapprox1.3249M.
\end{eqnarray}
We find that the radius of the ISCO for the spinning test particle is smaller than that of the non-spinning test particle in the Schwarzschild black hole background.

\begin{figure}[!htb]
    \subfigure[~$a=0, Q=0.1$ ]{
    \includegraphics[width=0.35\textwidth]{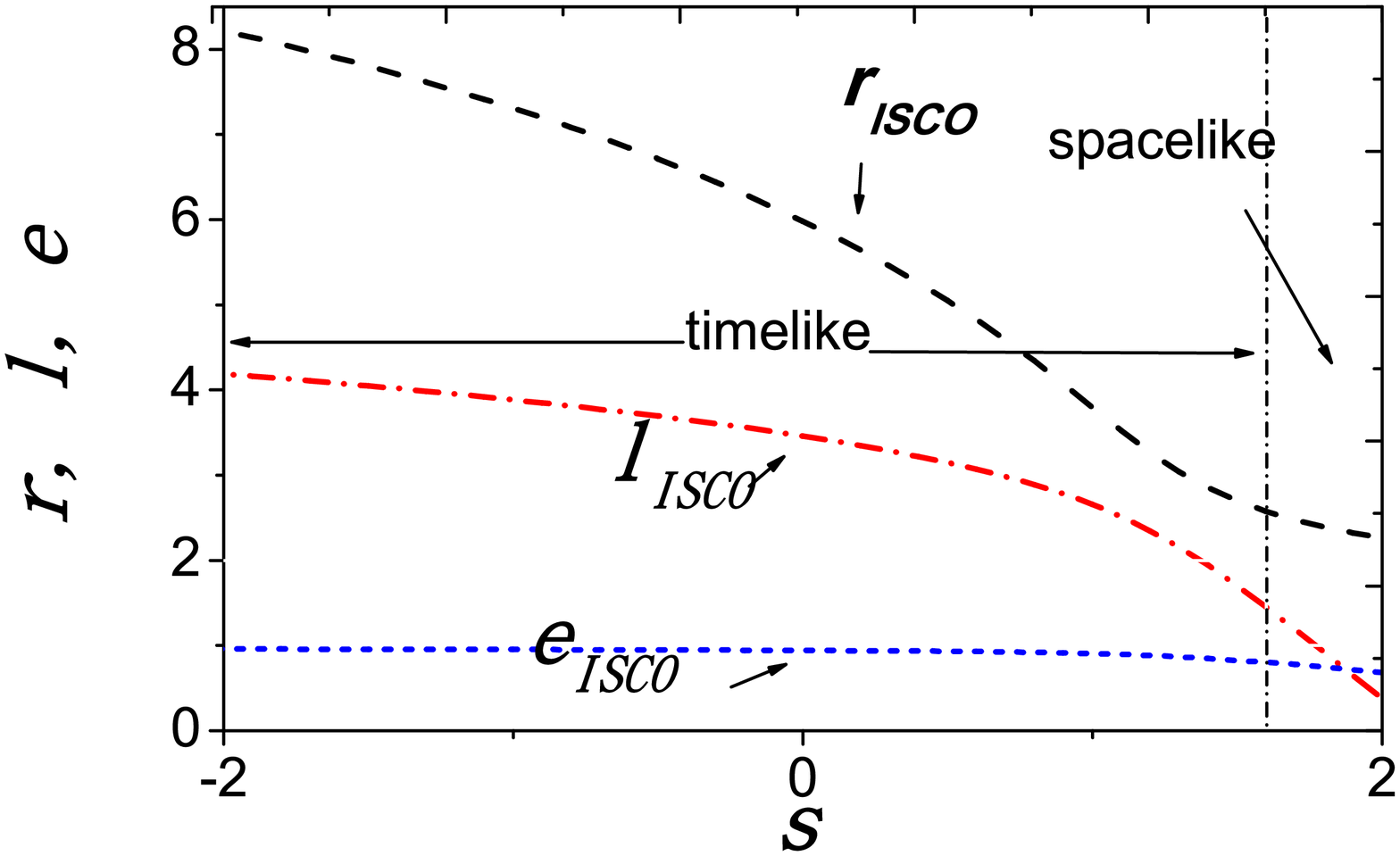}}
    \subfigure[~$a=0, Q=0.6$ ]{
    \includegraphics[width=0.35\textwidth]{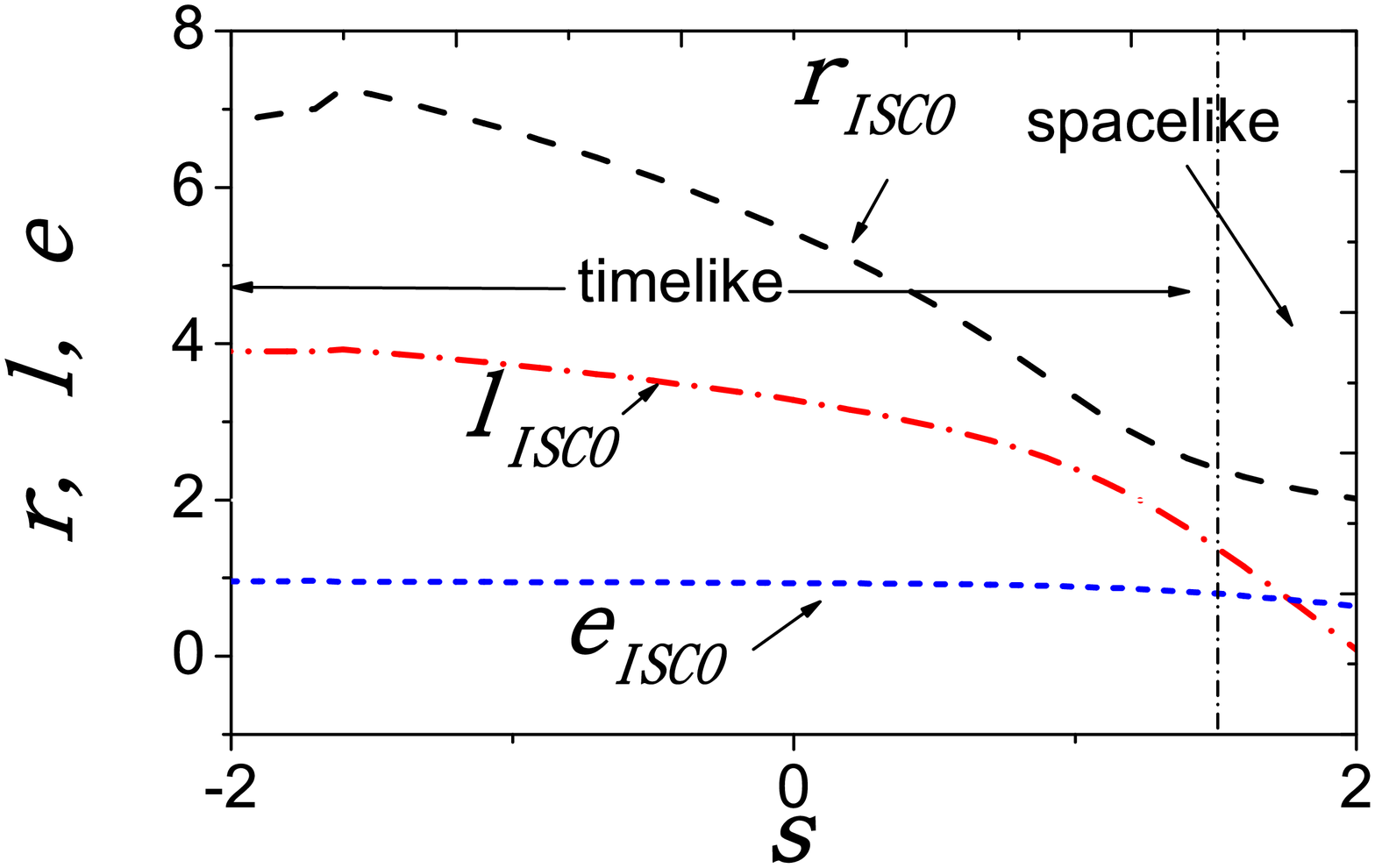}}
    \subfigure[~$a=0, Q=1$ ]{
    \includegraphics[width=0.35\textwidth]{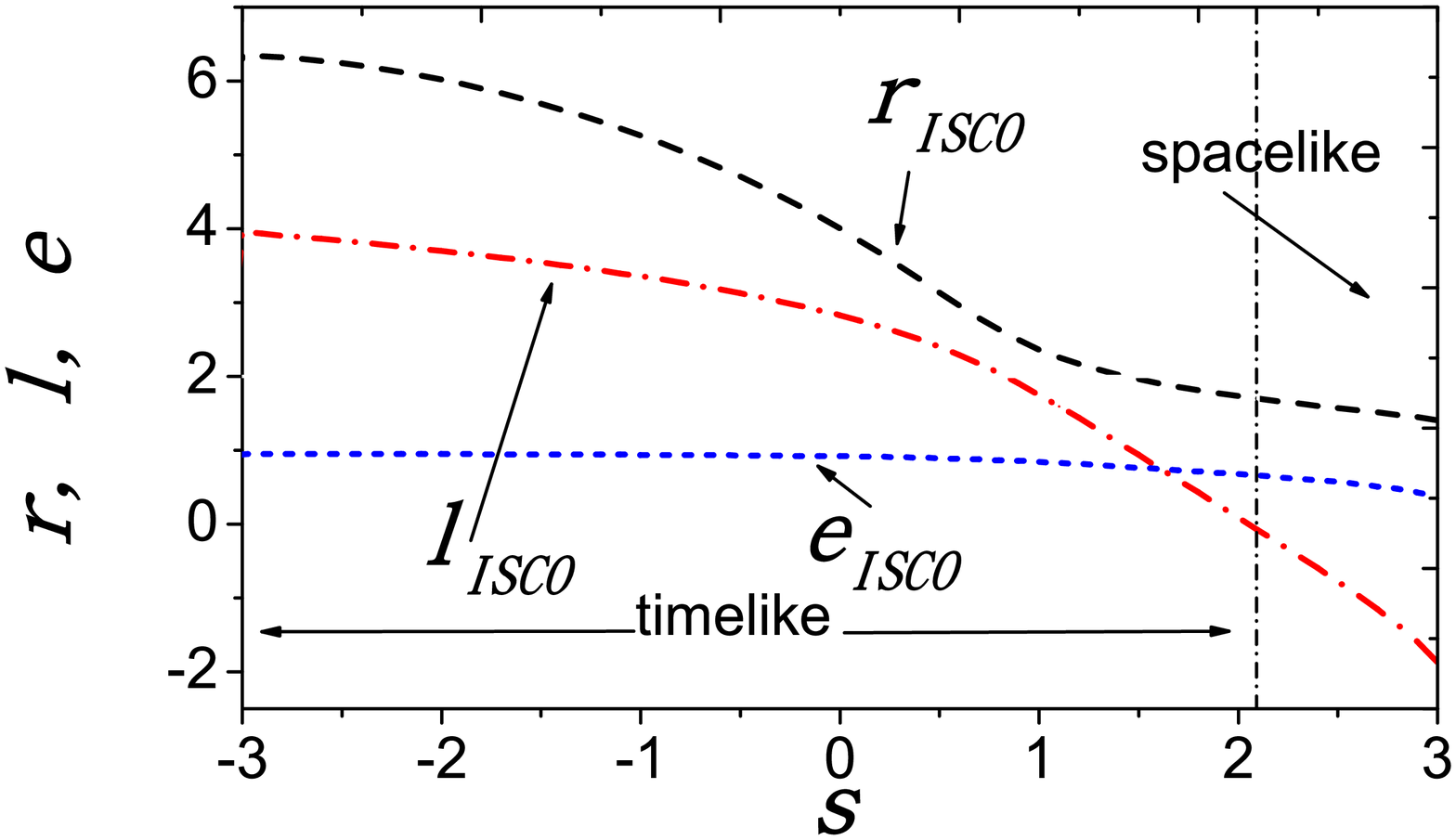}}
    \vskip -4mm \caption{Plots of the ISCO parameters of the spinning test particle as functions of the spin $s$ in the RN black hole background. The left side of the vertical line stands for that the orbit of the ISCO is timelike and physical, while the right side stands for that the orbit is spacelike and unphysical.}
    \label{rniscoparameter}
    \end{figure}

We know that the RN black hole is charged and its charge $Q$ will affect the effective potential \eqref{spinningeffective} of the test particle. The numerical results of the ISCO with different values of the charge $Q$ and spin $s$ are given in Fig.~\ref{rniscoparameter}.
For the charged black hole, we can see that the radius of the ISCO of the spinning test particle also decreases with the spin $s$, and some orbits are also superluminal. Note that the radius of the ISCO for the spinning test particle also decreases with the charge $Q$. The corresponding physical ISCO parameters of the spinning test particle in the extremal RN black hole background are
\begin{eqnarray}
\bar{s}^{\text{RN}}_{\text{ISCO}}&\thickapprox&2.1490M,~~~r^{\text{RN}}_{\text{ISCO}}\thickapprox1.6833M,~~~\nonumber\\
\bar{e}^{\text{RN}}_{\text{ISCO}}&\thickapprox&0.6474,~~~~~\,\,\, \bar{l}^{\text{RN}}_{\text{ISCO}}\thickapprox-0.1658 M.
\end{eqnarray}
Obviously, the radius of the ISCO of the spinning test particle in the charged black hole background is smaller than that in the background of the Schwarzschild black hole with the same mass $M$.

\begin{figure}[!htb]
    \subfigure[~$a=0.3, Q=0$ ]{
    \includegraphics[width=0.35\textwidth]{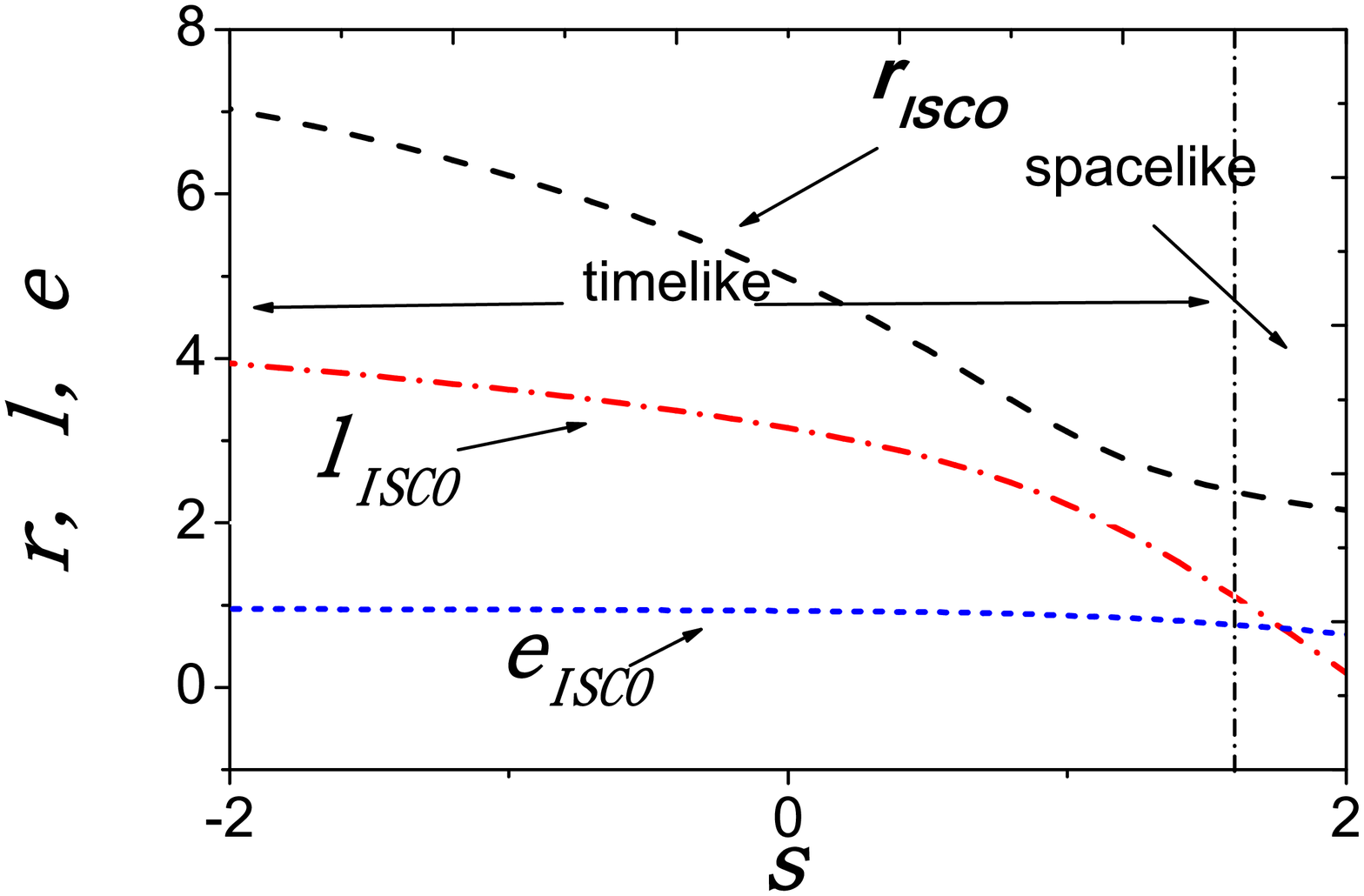}}
    \subfigure[~$a=0.8, Q=0$ ]{
    \includegraphics[width=0.35\textwidth]{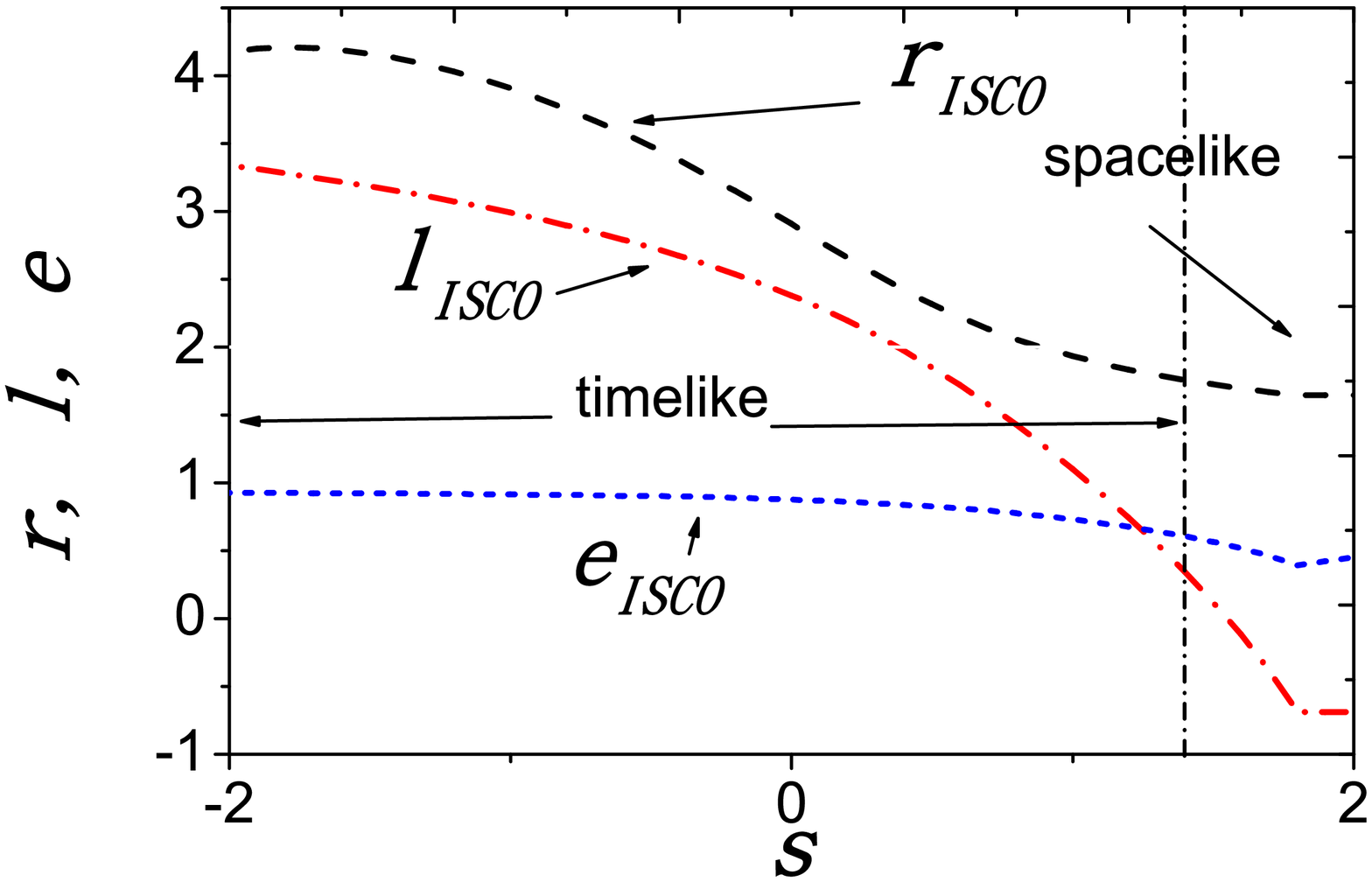}}
    \vskip -4mm \caption{Plots of the ISCO parameters of the spinning test particle with co-rotating orbit as functions of the spin $s$ in the Kerr black hole background. The left side of the vertical line stands for that the orbit of the ISCO is timelike and physical, while the right side stands for that the orbit is spacelike and unphysical.}
    \label{kerriscoparameter}
    \end{figure}

It is easy to know that the motion of a spinning test particle also depends on the spin of the black hole. The corresponding numerical results of the ISCO with co-rotating orbits in the Kerr black hole background are shown in Fig.~\ref{kerriscoparameter}. The ISCO of the spinning test particle with co-rotating orbit in the Kerr black hole background has the same behavior with the spin $s$ changing in the Schwarzschild and RN black hole cases, and the radius of the ISCO also decreases with the black hole spin $a$.

\begin{figure}[!htb]
    \subfigure[~$a=0.3, Q=0$ ]{
    \includegraphics[width=0.35\textwidth]{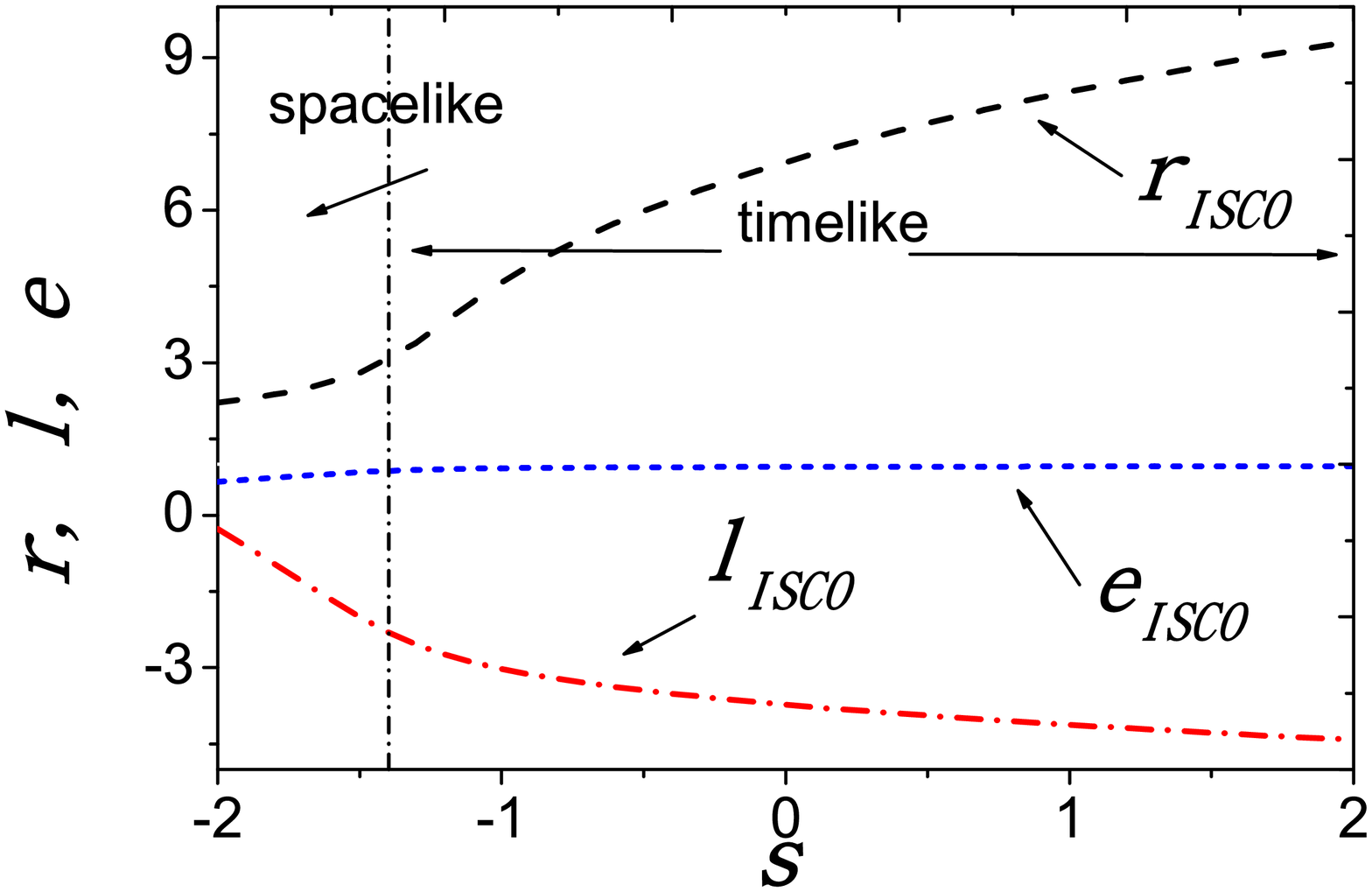}}
    \subfigure[~$a=0.8, Q=0$ ]{
    \includegraphics[width=0.35\textwidth]{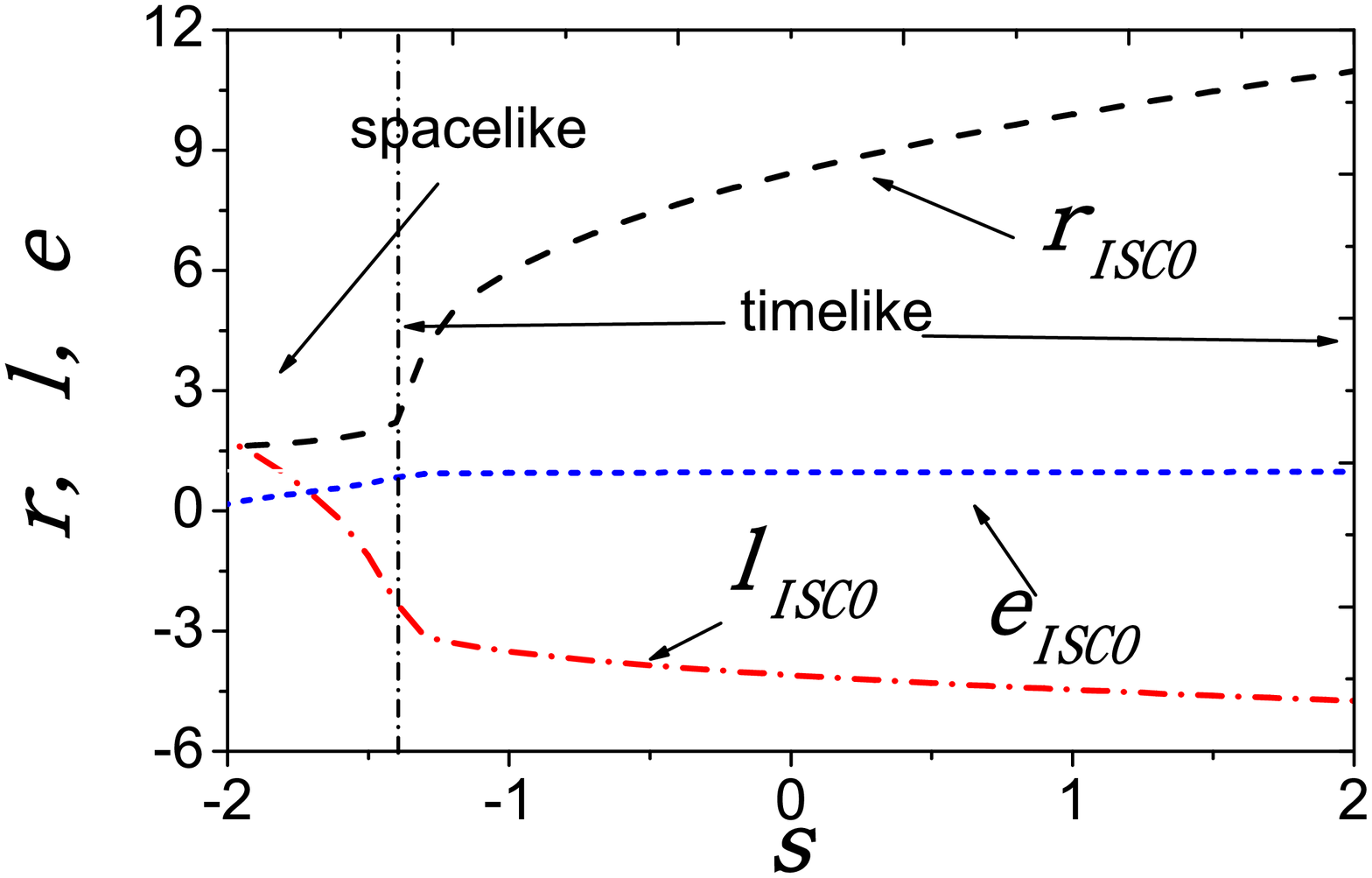}}
    \subfigure[~$a=1, Q=0$ ]{
    \includegraphics[width=0.35\textwidth]{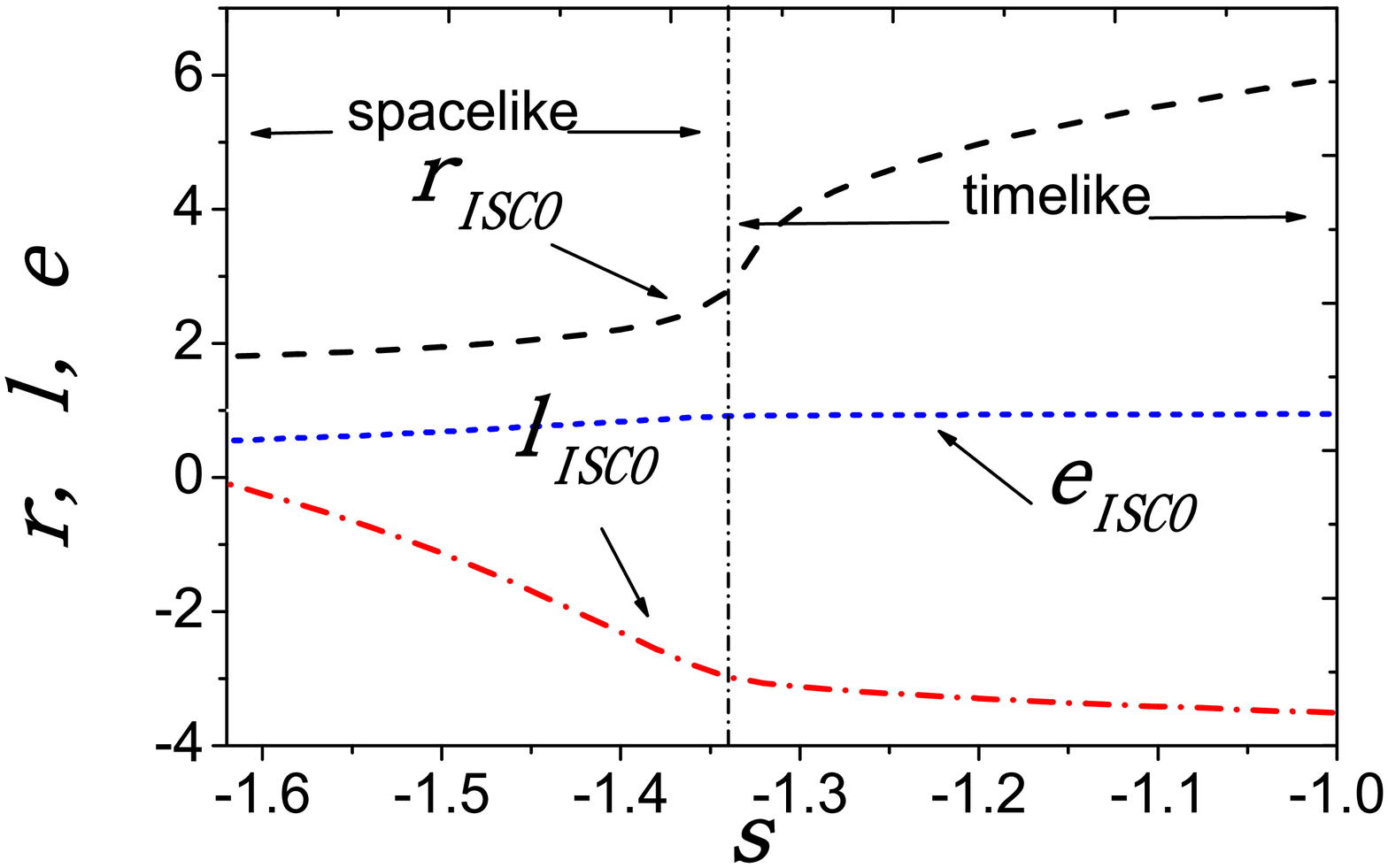}}
    \vskip -4mm \caption{Plots of the ISCO parameters of the spinning test particle with counter-rotating orbit as functions of the spin $s$ in the Kerr black hole background. The right side of the vertical line stands for that the ISCO is timelike and physical, while the left side stands for that the orbit is spacelike and unphysical.}
    \label{antikerriscoparameter}
    \end{figure}

The corresponding ISCO parameters for the counter-rotating or co-rotating orbits of the non-spinning test particle in the extremal Kerr black hole background are given in Eqs. \eqref{kerrisco1} and \eqref{kerrisco2}. Note that the radius of the ISCO with co-rotating orbit is
\begin{equation}
r_{\text{ISCO}}=M=r_h,
\end{equation}
which means that the radius of the ISCO with co-rotating orbit cannot decrease anymore in the extremal Kerr black hole background. While for the ISCO with counter-rotating orbit the corresponding radius can be smaller due to the existence of the spin $s$. The corresponding numerical results of the ISCO with counter-rotating orbit are shown in Fig.~\ref{antikerriscoparameter}.

The physical ISCO parameters of the spinning test particle with counter-rotating orbit in the extremal Kerr black hole background are given as follows
\begin{eqnarray}
\bar{s}^{\text{Kerr}}_{\text{ISCO}}&\thickapprox&-1.3200M,~~~~ r^{\text{Kerr}}_{\text{ISCO}}\thickapprox3.5890M,\nonumber\\
\bar{e}^{\text{Kerr}}_{\text{ISCO}}&\thickapprox&0.9237,~~~~~~~\,\,~~ \bar{l}^{\text{Kerr}}_{\text{ISCO}}\thickapprox-3.0712M.
\end{eqnarray}

\begin{figure}[!htb]
    \subfigure[~$a=0.5, Q=0.5$]{
    \includegraphics[width=0.35\textwidth]{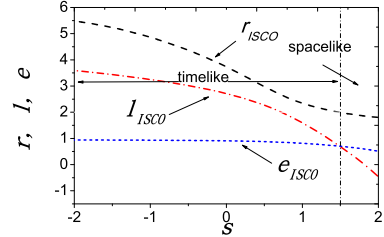}}
    \subfigure[~$a=0.5, Q=0.5$]{
    \includegraphics[width=0.35\textwidth]{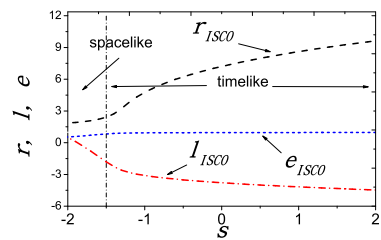}}
    \vskip -4mm \caption{Plots of the ISCO parameters of the spinning test particle with (a) co-rotating orbit and (b) counter-rotating orbit as functions of the spin $s$ in the KN black hole background. Only the left (right) side of the vertical line in upper (lower) figure stands for the timelike orbit.}
    \label{kniscoparameter}
    \end{figure}

We have shown that the spinning test particle can orbit with more smaller radius in the RN and Kerr black hole backgrounds than the case in Schwarzschild black hole with the same mass $M$. The numerical results of the ISCO in the KN black hole background are also given in Fig.~\ref{kniscoparameter}.

We can make a brief summary that the change of parameters for the black hole and test particle can yield the following results:

(1) For the physical ISCO of the spinning test particle in the Schwarzschild black hole background, the corresponding radius and angular momentum decrease with the spin $s$, which indicates that the spinning test particle can orbit with more smaller radius than the non-spinning test particle with stable circular orbit.

(2) For the physical ISCO of the spinning test particle in the RN black hole background, the corresponding radius and angular momentum also decrease with the spin $s$, and this behavior is the same as the Schwarzschild case. In addition to the effect resulted from the spin $s$, the corresponding radius and angular momentum of the ISCO also decrease with the charge of the black hole $Q$, and the ISCO in the charged black hole is smaller than the Schwarzschild case with the same mass $M$.

(3) For the physical ISCO of the spinning test particle with co-rotating orbit in the non-extremal Kerr black hole background, the corresponding radius and angular momentum also decrease with the spin $s$, which is consistent with the Schwarzschild and RN cases. The corresponding radius and angular momentum of the ISCO also decrease with the spin of the black hole $a$. We should note that the radius of the ISCO in the extremal Kerr black hole background with co-rotating orbit is the radius of the horizon $r_h$. So the ISCO in the extremal Kerr black with co-rotating orbit can not be decreased anymore. For the ISCO with counter-rotating orbit in the non-extremal Kerr black hole background, the radius of the ISCO decreases with the black hole spin $a$.

(4) For the physical ISCO of the spinning test particle in the KN hole background, the corresponding radius and angular momentum also decrease with the spin $s$. We have shown that the radius of the ISCO will decrease with the spin $a$ and charge $Q$ of the black hole. The most smallest radius of the ISCO in the KN black always appears in the case of the extremal KN black hole with spin $a=1$ (extremal Kerr black hole).

\section{Summary and conclusion}\label{Conclusion}

In this paper, we have numerically investigated the ISCO of a spinning test particle in the Schwarzschild black hole, RN black hole, Kerr black hole, and KN black hole backgrounds. We used Eqs. \eqref{equationmotion1} and \eqref{equationmotion2} to describe the motion of the spinning test particle in curved spacetime, while the four-velocity of the spinning test particle can be transformed from timelike into spacelike due to the four-velocity vector $u^\mu$ and the conjugate momentum $P^\mu$ are not parallel. So we gave the fourth condition \eqref{velocitysquare} by considering the superluminal constraint to obtain the physical ISCO of the spinning test particle in various black hole backgrounds. We numerically gave the relations between the ISCO parameters and the spin $s$ and showed that a spinning test particle can orbit in more smaller circular orbit than a non-spinning test particle. The radius of the ISCO for a spinning test particle was also affected by the charge and spin of the black hole, and the radius of the ISCO in the RN and Kerr black hole backgrounds are smaller than the case in Schwarzschild black hole. Although the radius of the ISCO decreases with the spin of the test particle, we should note that the radius of the ISCO with co-rotating orbit in an extremal Kerr black can not decrease any more because the corresponding radius is the horizon of the extremal Kerr black hole.

\acknowledgments{
This work was supported in part by the National Natural Science Foundation of China (Grants Nos. 11522541, 11375075, 11675064)}, and the Fundamental Research Funds for the Central Universities (Grants Nos. lzujbky-2017-it68, lzujbky-2016-115, and lzujbky-2016-k04).

\end{document}